\documentstyle[12pt,epsf,psfig]{article}

\def\beq{\begin{equation}}
\def\eeq{\end{equation}}
\def\beqa{\begin{eqnarray}}
\def\eeqa{\end{eqnarray}}
\def\a {{\rm f}}
\def\e{r}
\def\g{\xi}
\def\w{\rho}

\begin{document}

\begin{flushright}
ITP-SB-98-70
\end{flushright}

\begin{center}{\bf\Large\sc Dijet Rapidity Gaps in Photoproduction 
from Perturbative QCD}
\vglue 1.2cm
\begin{sc}
Gianluca Oderda \\
\vglue 0.5cm
\end{sc}
{\it Institute for Theoretical Physics \\
State University of New York at Stony Brook \\
Stony Brook, NY 11794-3840, USA}\\
\end{center}
\vglue 1cm
\begin{abstract}
By defining dijet rapidity gap events  according
to interjet energy flow, 
we treat the photoproduction cross section 
of two high transverse momentum jets with a large intermediate
rapidity region as a factorizable quantity in perturbative QCD.
We show that logarithms of soft gluon energy in the interjet region
can be resummed to all
orders in perturbation theory. The resummed cross section depends on 
the eigenvalues of a set of soft anomalous dimension matrices, specific
to each underlying partonic process, and  on the decomposition of 
the scattering
according to the possible patterns of hard color flow. We present a detailed
discussion of both. Finally, we evaluate numerically the gap cross section and gap fraction
and compare the results with ZEUS data. 
In the limit of low gap energy,
good agreement with experiment is obtained.
\end{abstract}

\centerline{PACS Nos.: 12.38.Aw, 12.38.Cy, 13.85.-t, 13.87.-a} 

\newpage

\section{Introduction}
In a recent paper \cite{StOd1}
we have presented an explanation, from perturbative QCD,
of the dijet rapidity gap events experimentally observed 
in $p \bar{p}$ scattering at the Fermilab Tevatron \cite{D0CDF}.
These events consist of a pair of jets
produced at very large momentum transfer, 
and separated by a wide empty region,
where barely any particle multiplicity is measured.
They were originally predicted from the exchange of
at least two gluons in a color singlet configuration,
which avoids color recombination between the jets and
associated interjet hadronization \cite{Bjork}.
A quantitative formalism to describe gap events,
however, has been lacking in the literature.

In Ref.\ \cite{StOd1} we have discussed 
the dependence of the dijet cross section on the energy flow into
the central region, clearly related to the observed particle multiplicity.
We have shown how to factorize the partonic cross section 
into a hard scattering function, accounting for the dynamics of 
the highly virtual quanta exchanged in the scattering, and a soft function,
describing the soft 
radiation emitted into the interjet region.
Both functions are defined as matrices in the space
of the possible color flows at the hard scattering,
formalizing the idea,
already expressed by other authors, that 
the color structure of the short-distance partonic scattering  
is not uniquely defined \cite{BuchHeb,eboli}.
The soft function contains, at each order in perturbation theory,
logarithms of the interjet radiation energy, which can
be resummed, to give the dependence of the dijet cross section 
on the interjet energy flow.
The resummation is driven by the eigenvalues and eigenvectors of 
a soft anomalous dimension matrix, defined in the space of color flows.
In Ref.\ \cite{StOd1}, we have performed the  analysis of this matrix
for $q \bar{q} \rightarrow q \bar{q}$ scattering, which
is the relevant partonic process in a valence
quark approximation for $p \bar{p}$ collisions. 
The result indicates that,
in the limit of a very large interjet region 
(corresponding to a very large parton center of mass energy,
compared to the momentum transfer),
the color singlet component dominates, thus merging our
picture, in this asymptotic configuration, with approaches
to the problem based on the Regge limit of QCD \cite{DDucaT} and on the
idea of color singlet dominance
\cite{Bjork,zepp}. 
We have already pointed out that a full treatment of the Tevatron 
dijet gap cross section
requires us to include the contribution of processes involving gluons, which we
postpone, for that specific problem, to forthcoming work \cite{StOd}. 

In this paper, we will apply our method to analyze   
the photoproduction of two high transverse momentum jets with 
a large intermediate rapidity gap. 
Events of this kind have been reported in $e^{+} p$ collisions
at HERA \cite{ZEUS}.
Here, the quantity of interest is the gap fraction, the ratio
of the number of dijet events with a large interjet gap, to
the total number of dijet events. 
The ZEUS experiment shows that, binning the gap events according
to the width of the interjet empty region,  
the gap fraction, after an initial decay, reaches an asymptotic plateau.
It has been shown in Refs.\ \cite{Bjork, DDucaT} that the 
fraction of gap events from the exchange of a
color octet, reggeized gluon falls off exponentially with increasing
gap width.
The left-over asymptotic excess has been interpreted as the fraction of the color singlet
exchange of a hard BFKL pomeron \cite{DDucaT}. 
We emphasize here that the BFKL approach 
is complementary to ours, because it deals with
the resummation of gluonic radiation, which, in our language, would be 
described as hard (see Sec.\ \ref{results} below).

Applying the methods of Ref.\ \cite{StOd1} to this problem is especially
interesting, because, while the data for the Tevatron gaps show
the dependence of the cross section on the interjet energy at
fixed gap width, here we will explore the opposite dependence, variable
gap width at fixed energy.
We will show that the factorization and resummation of the soft gluonic energy
emitted into the gap leads precisely to the behavior of the gap fraction
observed by the experiment.
In addition, photoproduction is an ideal process in which 
to analyze the contribution 
of gluons to gap cross sections.  
Theoretical studies of jet photoproduction have been pursued since the
early 80's \cite{photold,Aur,FRGS,Drees,KK,phrev} and are a very active area of research.
The dominant partonic mechanisms  in photoproduction 
are the direct and  resolved scatterings: in the former case the photon  from the incoming lepton
interacts directly with the quark or gluon from the proton, and in the latter it  fluctuates first into 
a hadronic state of low virtuality, acting as a source of partons, which then scatter off partons from the 
proton. 
The precise experimental determination of the partonic content of the photon is still
an open problem, partly because it requires isolating
the resolved component of the scattering \cite{photold,ZEUS2}.
However, existing data indicate that, especially at low values of 
partonic $x$, 
the dominant component of the photon is gluonic \cite{FRGS}.
Correspondingly, we will see that for the resolved contribution, the 
scattering of a gluon from the photon with a quark from the proton
is dominant in the kinematical region of interest.
Also the contribution from the resolved reaction $g+g \rightarrow g+g$ has to be taken into account,
although it turns out to be much less important in this case, 
because of the high transverse energy cuts experimentally imposed on the 
jets.

One last important remark has to be made, about the issue of survival \cite{Bjork, GLM}. 
In dijet gap events, the survival probability is limited by the probability
of no radiation inside the gap  from the interaction of spectator partons.
Such probability is estimated to be of the order of $10 \%$  
in $p\bar{p}$ scattering, if we require a truly empty interjet region \cite{Bjork,GLM}.
Presumably, for $ep$ reactions the number of spectator quarks
and gluons is reduced with respect to $p\bar{p}$ scattering,
because of the partly leptonic nature of the initial state.
Therefore our resummed formula, which accounts for the amount of perturbative survival
from soft gluon dynamics, should be less sensitive to these effects.

Throughout the paper, we will work in close correspondence 
with the ZEUS experimental
configuration described in Ref.\ \cite{ZEUS}.
In Sec.\  \ref{gapfacto}, we will discuss the kinematics
of the problem, define the dijet gap cross section,
and review its factorization properties. 
In Sec.\  \ref{fraction}, we will introduce
the gap fraction, as the ratio of the dijet gap cross section to the overall dijet cross section.
We will also identify the partonic reactions giving the largest contributions to both the gap
and the overall cross section. 
For each of these reactions, in Sec.\  \ref{hardsoft}, 
we will present the explicit decomposition into hard and soft parts. 
In Sec.\  \ref{anodim}, we will present the process-dependent
soft anomalous dimension matrices, their eigenvalues and eigenvectors, which govern
the soft dynamics of the scattering.
Finally, in Sec.\  \ref{results}, we will give numerical results for
the overall dijet cross section, the dijet gap cross section and the
gap fraction.  We will draw a comparison  with the experimental results of 
Ref.\ \cite{ZEUS}, and present our conclusions.

\section{Dijet Gap Cross Sections in Photoproduction}
\label{gapfacto}
\subsection{Definitions}
In this section we will introduce dijet gap cross sections in photoproduction.
We first recall the definition of particle pseudorapidity,
\beq
\eta=\ln \left( \cot \left(\theta \over 2 \right) \right),
\label{pseudoi}
\eeq
where $\theta$ is the angle of a particle momentum with respect
to a fixed direction, typically the beam direction.
We will consider positron-proton scattering
\beq
e^{+}(p_A)+p(p_B) \longrightarrow e^{+}(p'_A)+J_1(p_1)+J_2(p_2)+X_{\rm{gap}},
\label{process}
\eeq
for the production of two jets at fixed pseudorapidity difference, $\Delta \eta=\eta_1-\eta_2$.
We sum inclusively over final states, while measuring the (soft) energy flow into the intermediate
region between the jets.
The boost-invariant pseudorapidity difference $\Delta \eta$ 
fixes the partonic scattering angle $\hat{\theta}$
through the formula
\beq
\frac{\Delta \eta}{2}=\ln \left( \cot \left( \frac{\hat{\theta}}{2} \right) \right).
\label{deltaeta}
\eeq 
Following Ref.\ \cite{ZEUS}, the 
jets are defined by cones of radius $R=1.0$ in the 
$\eta$-$\phi$ plane mapping the ZEUS calorimeter detector.
This constrains the interjet region to have (pseudo)rapidity width
\beq
\Delta y=\Delta \eta-2R.
\label{dydeta}
\eeq 

The low-virtuality photons exchanged at the electromagnetic vertex of the interaction
can be thought of as real particles of energy $E_{\gamma}=yE_e$, whose spectrum is given by the Weisz\"acker-Williams
formula \cite{WWil,MaRiNa}
\beq
F_{\gamma/e}(y)=\frac{\alpha_{\rm{em}}}{2\pi}\left\{\frac{1+(1-y)^2}{y} \ln
\left( \frac{Q^2_{\rm{max}}(1-y)}{m_e^2 y^2}\right)+2 m_e^2 y \left[
\frac{1}{Q^2_{\rm{max}}}-\frac{(1-y)}{m_e^2 y^2} \right] \right\}.
\label{weiszwi}
\eeq
Here $\alpha_{\rm{em}}$ is the electromagnetic coupling, $m_e$ is the electron mass,
and $Q^2_{\rm{max}}$ is the maximum photon virtuality, determined from the (anti) tagging
conditions of the experiment. From Ref.\ \cite{ZEUS} we take the value $Q^2_{\rm{max}}=4 \rm{GeV}^2$. 
The generic differential cross section for electron-proton scattering can then be viewed as a 
convolution of the photon distribution in the electron and the photon-proton cross 
section:
\beq
d\sigma_{ep}(S)=\int_{y_{\rm{min}}}^{y_{\rm{max}}}{dy F_{\gamma/e}(y) \, d\sigma_{\gamma p}(S_{\gamma p})}, 
\label{gencs}
\eeq
where $S$ is the center of mass energy squared of the electron-proton system, and where $S_{\gamma p}=yS$
is its analog for the photon-proton system. 
We use for $y_{\rm{min}}$ and $y_{\rm{max}}$ the values given in Ref.\ \cite{ZEUS}, $y_{\rm{min}}=0.2$
and $y_{\rm{max}}=0.8$.
\subsection{The factorized cross section}
For our specific case,
the inclusive cross section for dijet events with transverse energy greater than $E_T$,
rapidity difference
$\Delta \eta$, and energy flow less than $Q_c$ in the intermediate region, of rapidity width $\Delta y$,  
can be written as follows \cite{photold,Aur,FRGS,Drees,KK}:  
\beqa
\frac{d\sigma_{ep}}{d\Delta\eta }\left(Q_c,S,E_{T},\Delta y\right) 
&=&\int_{y_{\rm{min}}}^{y_{\rm{max}}} dy F_{\gamma/e}(y) \nonumber \\
&& \hspace{-5mm} \times \left[ \frac{d\sigma^{\rm{dir}}_{\gamma p}}
{d\Delta\eta }\left(Q_c,S_{\gamma p},E_{T},\Delta y\right)+
 \frac{d\sigma^{\rm{res}}_{\gamma p}}{d\Delta\eta}\left(Q_c,S_{\gamma p},E_{T},\Delta y\right)\right].
\label{factcs1}
\eeqa
Here $d\sigma^{\rm{dir}}_{\gamma p}/{d\Delta\eta}$ and $d\sigma^{\rm{res}}_{\gamma p}/{d\Delta\eta }$,
are the direct and resolved contributions to the cross section respectively.
In the former case the low-virtuality photon interacts directly with the parton from the proton, while
in the latter, as mentioned above, it acts itself as a source of quarks and gluons, 
which then scatter off the partons from the proton.
The two cross sections can be written in factorized form as \cite{photold,Aur,FRGS,Drees,KK}
\beqa
\frac{d\sigma^{\rm{dir}}_{\gamma p}}{d\Delta\eta }\left(Q_c,S_{\gamma p},E_{T},\Delta y\right)
&=&\sum_{f_p,f_1,f_2} \,
\int_{R_{x_p}} dx_p \,
\phi_{f_p/p}(x_p,-\hat{t})  \nonumber \\ 
&&  \times  \,
 \frac{d\hat{\sigma}^{(\gamma \a)}}
{d\Delta\eta} 
\left(Q_c,\hat{t},\hat{s},\eta_{JJ},\Delta y,\alpha_s(\hat{t})\right)\, ,
\label{crosdir}
\eeqa
and 
\beqa
\frac{d\sigma^{\rm{res}}_{\gamma p}}{d\Delta\eta }\left(Q_c,S_{\gamma p},E_{T},\Delta y\right)
&=& \sum_{f_{\gamma},f_p,f_1,f_2} \,
\int_{R_{x_{\gamma}}} dx_{\gamma} \int_{R_{x_p}} dx_p \,
\phi_{f_{\gamma}/{\gamma}}(x_{\gamma},-\hat{t}) \, \phi_{f_p/p}(x_p,-\hat{t})  \nonumber \\ 
&& \hspace{9mm} \times 
\, \frac{d\hat{\sigma}^{(\a)}}
{d\Delta\eta} 
\left(Q_c,\hat{t},\hat{s},\eta_{JJ},\Delta y,\alpha_s(\hat{t})\right) \, . \nonumber \\
\label{crosres}
\eeqa
In these formulas $\phi_{f_{\gamma}/{\gamma}}$ and $\phi_{f_p/p}$ are parton distributions in
the photon and proton respectively, evaluated at the scale $-\hat{t}$, the dijet momentum transfer,
which is related to the  partonic center of mass energy squared, $\hat{s}$, and the dijet rapidity difference,
$\Delta \eta$, according to the formula 
$\hat{t}=-\frac{\hat{s}}{2} \left(1-\tanh(\frac{\Delta \eta}{2}) \right)$.
The integration regions for the partonic fractions $x_p$ and $x_{\gamma}$
are denoted by $R_{x_p}$ and $R_{x_{\gamma}}$ respectively. 
${d\hat{\sigma}^{(\gamma \a)}}/{d\Delta\eta }$ and ${d\hat{\sigma}^{(\a)}}/{d\Delta\eta }$
are hard scattering functions, starting from the lowest order Born cross section.
The index $\a$ ($\gamma \a$) denotes the partonic process $f_{\gamma}+f_p \rightarrow f_1+f_2$
($\gamma+f_p \rightarrow f_1+f_2$).
The detector geometry constrains the phase space for the dijet total pseudorapidity, $\eta_{JJ}=
(\eta_{J_1}+\eta_{J_2})/{2}$, with $|\eta_{JJ}|<0.75$ \cite{ZEUS}.
Similarly, the lower bound on the transverse energy of the jets, $E_T$, and the dijet pseudorapidity
determine the phase space for   
the partonic center of mass energy squared $\hat{s}$, with
$4{E_T}^2\cosh^2 \left( {\Delta \eta} \over 2 \right)< \hat{s} < S_{\gamma p} \exp (2\eta_{JJ}) \, yE_e/E_p $.
\subsection{Refactorization of the partonic scattering}
\label{refacto}
In this section we will review some of the arguments already presented in 
Refs.\ \cite{StOd1,BottsSt,KOS1,KOS2}, to show
how it is possible to perform a further factorization on the partonic scattering functions,
${d\hat{\sigma}^{(\gamma \a)}}/{d\Delta\eta }$ and ${d\hat{\sigma}^{(\a)}}/{d\Delta\eta}$,
of Eqs.\ (\ref{crosdir}) and (\ref{crosres}). 
The underlying argument is that in the partonic scattering 
the soft gluon emission decouples from the dynamics of the hard scattering,
and can be approximated by an effective cross section, in which each of the partons 
is treated as a recoilless source of gluonic radiation.
Formally, this is equivalent to replacing each parton with a path-ordered exponential of the gluon field
(eikonal or Wilson line) in the proper representation of $SU(3)$, the fundamental representation
$3$ ($3^*$) for quarks (antiquarks), and the adjoint representation for gluons. 
In this way the hard amplitude is replaced by a sum of eikonal operators, $w_I(x)_{\{c_k\}}$,
depending on color tensors $c_I$,
which account for the color flow at the hard scattering, times short-distance coefficient functions.
The effective dimensionless eikonal cross section 
( in the following $\a$ will refer to both direct and resolved processes)
can be written as
\beqa
\hat{\sigma}_{LI}^{(\a, {\rm{eik}})}\left(\frac{Q_c}{\mu},\Delta y\right)
&=&
\sum_{\xi}\, \theta\left(Q_c-E_c(\xi)\right)
\nonumber\\
&\ & \hspace{-30mm} \times 
\langle0|{\bar T}\left[ \left(w_L(0)\right){}^{\dagger}_{\{b_i\}}\right]|\xi{\rangle}
{\langle}\xi|T\left[w_I(0)_{\{b_i\}}\right]|0 \rangle \, , \nonumber \\
\label{eq:eikcs}
\eeqa
where we sum over all the final states subject to the constraint of having 
energy less than $Q_c$ in the  interjet region of rapidity $\Delta y$, while
cutting off all the remaining integrals at the ultraviolet scale $\mu$.
This makes the eikonal cross section free from potential collinear singularities associated with gluon emission
from the Wilson lines \cite{StOd,KOS1,KOS2}. 
The Latin indexes $I$ and $L$ refer to the color structures of the amplitude and of its
complex conjugate.
At tree level, with no soft gluons, the above formula reduces to
the square of eikonal vertices, with matrix elements given by  traces
of the color tensors in the amplitudes.

In these terms, the partonic scattering function can now be factorized
into the product, in the space of color tensors, of a hard scattering matrix
\footnote{The normalization of the dimensionless hard scattering matrix, $H^{(\a)}_{IL}$,
can be found in Eq.\ (\ref{analcs}) below. 
It differs slightly  from the one used in Ref.\ \cite{StOd1}.}   
, $H^{(\a)}_{IL}$, accounting for the quanta of
high virtuality exchanged in the scattering, and a soft matrix, $S^{(\a)}_{LI}$,
\beqa
\frac{d\hat{\sigma}_{\gamma p}
^{(\a)}}{d\Delta \eta }\left(Q_c,\hat{t},\hat{s},\eta_{JJ},\Delta y,\alpha_s(\hat{t})\right)
&=&\, \frac{\pi}{2\hat{s}} \, \left( 2 \cosh^2 \left( \frac{\Delta \eta}{2} \right)\right)^{-1}  \nonumber \\
&& \hspace{-30mm} \times H^{(\a)}_{IL}\left(\sqrt{-\hat{t}},
\sqrt{\hat{s}},\mu,\alpha_s(\mu^2) \right) \, 
 S^{(\a)}_{LI} \left(\Delta y,\frac{Q_c}{\mu}\right)\,,
\label{factor}
\eeqa
where we follow the convention of the sum over repeated indices. 
This factorization holds to leading power in $\frac{\Lambda}{Q_c}$,
with $\Lambda$ the QCD scale parameter.
We can identify a hard scale, $\sqrt{-\hat{t}}$,  a soft scale, $Q_c$,
and a new factorization scale, $\mu$. The soft matrix, $S_{LI}$, precisely coincides
with the gauge invariant eikonal cross section of Eq.\ (\ref{eq:eikcs}),
\beq
S^{(\a)}_{LI} \left( \frac{Q_c}{\mu},
\Delta y\right)= \hat{\sigma}_{LI}^{(\a, {\rm{eik}})}\left(\frac{Q_c}{\mu},\Delta y\right)\,.
\label{relSeikcs}
\eeq
In general, corrections to the factorized expression in Eq.\ (\ref{factor}) are expected from 
three-jet final states, 
but, since the cost of adding an extra jet to the final state is at least one power of $\alpha_s(\hat{t})$,
they will be suppressed.
\subsection{Evolution of the soft matrix in color space}
\label{evosec}
The left-hand side of the factorized expression, Eq.\ (\ref{factor}), is independent 
of $\mu$. This means that the $\mu$-dependence of the two matrices, $H^{(\a)}_{IL}$ and  $S^{(\a)}_{LI}$, must
cancel in the product. 
We thus derive for  $S^{(\a)}_{LI}$ the evolution equation \cite{StOd1}
\beq 
\left(\mu\frac{\partial}{\partial\mu}+\beta(g)\frac{\partial}{{\partial}g}
\right)S^{(\a)}_{LI}=
-(\Gamma^{(\a)}_S)^{\dagger}_{LB}S^{(\a)}_{BI}
-S^{(\a)}_{LA}(\Gamma^{(\a)}_S)_{AI}\, ,
\label{eq:resoft}
\eeq
where $\Gamma^{(\a)}_S(\alpha_s)$ is a process-dependent soft anomalous dimension
matrix. Solving this equation will enable us to resum all the leading logarithms of the soft scale $Q_c$.  
It is convenient to treat Eq.\ (\ref{eq:resoft}) in a basis which diagonalizes
$\Gamma^{(\a)}_S(\alpha_s)$.  Following
Ref.\ \cite{KOS2}, we denote by Greek indexes the basis of the  eigenvectors of $\Gamma^{(\a)}_S(\alpha_s)$,
$\{|e^{(\a)}_\beta \rangle\}$, corresponding to the eigenvalues $\lambda^{(\a)}_{\beta}$.
We then transform Eq.\ (\ref{factor}) to this basis and solve 
the evolution equation (\ref{eq:resoft}) for $S^{(\a)}$, by integrating with respect to $\mu$ between
the soft scale $Q_c$ and the hard scale $\sqrt{-\hat{t}}$, to get:
\beqa
\frac{d\hat{\sigma}_{\gamma p}
^{(\a)}}{d \Delta \eta }\left(Q_c, \hat{s},\hat{t},\eta_{JJ},\Delta y,\alpha_s(\hat t)\right)
&=&\, \frac{\pi}{2\hat{s}} \, \left( 2 \cosh^2 \left( \frac{\Delta \eta}{2} \right)\right)^{-1} 
\nonumber \\
&& \hspace{-40mm} \times \sum_{\alpha, \beta} H^{(\a,1)}_{\beta \alpha}\left( \Delta y,\sqrt{\hat{s}},\sqrt{-\hat{t}},
\alpha_s\left(\hat{t}\right) \right) 
 S^{{(\a,0)}}_{\alpha \beta} ( \Delta y ) \,  \nonumber \\
&& \hspace{-40mm} \times \, 
\left[\ln\left({Q_c\over \Lambda}\right)\right]^{E^{(\a)}_{\alpha\beta}}\;  
\left[ \ln \left( {\sqrt{-\hat{t}}\over \Lambda}\right)
\right]^{-E^{(\a)}_{\alpha\beta}} . \nonumber\\
\label{factor2}
\eeqa
The double differential cross section of Ref.\ \cite{StOd1}, ${d^2\hat{\sigma}_{\gamma p}
^{(\a)}}/{d \Delta \eta dQ_c}$, giving the distribution of dijet events
as a function of interjet radiation 
and rapidity interval, 
is obtained by differentiation with respect to $Q_c$.
In the above formula  the kinematical  cuts on the jets require the minimum value 
of $\sqrt{-\hat{t}}$ to be 
exactly $E_T$ ($E_T=5 {\rm{GeV}}$ in Ref.\ \cite{ZEUS}), still 
much larger than $\Lambda$. 
The exponents  $E^{(\a)}_{\alpha \beta}$ are given by
\beqa
&&E^{(\a)}_{\alpha\beta}\left(\Delta \eta,\Delta y \right)
=\frac{2\pi}{\beta_1}\, \left[{\hat{\lambda}^{(\a) \, *}_{\alpha}} \left(\Delta \eta,
\Delta y \right) +\hat{ \lambda}^{(\a)}_{\beta} \left( \Delta \eta,
\Delta y \right) \right]\, ,
\label{expon}
\eeqa
where $\beta_1$ is the first coefficient in the expansion of the QCD
$\beta$-function, $\beta_1=\frac{11}{3}N_c-\frac{2}{3}n_f$,
and where we define $\hat{ \lambda}^{(\a)}_{\beta}$ by 
$\lambda^{(\a)}_{\beta}=\alpha_s \hat \lambda^{(\a)}_\beta+\cdots$.
Observe that we have explicitly mantained a distinction between the dependence on 
$\Delta \eta$ and $\Delta y$, in spite of the fact that for this specific problem
they are related through the identity in Eq.\ (\ref{dydeta}). 
In fact $\Delta \eta$ can be seen as a parameter of the hard scattering (related to the 
scattering angle), whereas $\Delta y$,
at fixed $\Delta \eta$, is a measure of the width of the interjet region
and, as such, has a geometrical interpretation.  
The matrix $S^{{(\a,0)}}_{\alpha \beta}$ in Eq.\ (\ref{factor2}) 
is obtained by transforming to the new basis the zeroth order $S^{{(\a,0)}}_{LI}$ of 
Eq.\ (\ref{factor}), which, as mentioned above, is just a
set of color traces.
The change of basis goes through the matrix 
${\left( (R^{(\a)})^{-1} \right)}_{K \beta} \equiv \left( {e^{(\a)}_{\beta}}
\right)_K$, according to the formula \cite{KOS2}
\beq
S^{(\a,0)}_{\alpha \beta} \equiv 
\left[ \left( (R^{(\a)})^{-1}\right)^\dagger \right]_{\alpha M}
S^{(\a,0)}_{MN}
{\left( (R^{(\a)})^{-1} \right)}_{N \beta}\, .
\label{eq:newbasS}
\eeq
Analogously, we take for the $H^{(\a,1)}_{IL}$'s the squares of
the Born-level amplitudes, represented 
in the original color basis.
In the diagonal basis, each hard matrix $H^{(\a,1)}_{IL}$ becomes   
$H^{(\a,1)}_{\beta\alpha}$, defined by
\beq
H^{(\a,1)}_{\beta\alpha}={\left( R^{(\a)} \right)}_{ \beta L} \, 
H^{(\a,1)}_{LK}
\, {\left( R^{(\a)}{}^{\dagger} \right)}_{K \alpha}.
\label{newbasH}
\eeq 
Observe that  $S^{(\a,0)}$ and  $H^{(\a,1)}$ 
both acquire a $\Delta y$-dependence through the change of basis.

The relevant 
$\Delta \eta$-dependence in the right hand side of
Eq.\ (\ref{factor2}) is contained  
in the  exponents $E^{(\a)}_{\alpha\beta}$, Eq.\ (\ref{expon}).
We will see below that, for all the partonic processes, 
the real parts of the eigenvalues of $\Gamma^{(\a)}_S(\alpha_s)$
in the exponents 
are positive definite functions, with boundary 
value ${\rm{Re}}\left(\hat \lambda^{(\a)}_\beta\right)=0$ 
at $\Delta \eta=2$ ($\Delta y=0$). 
Then, since $Q_c < \sqrt{-\hat{t}}$, the last line in 
Eq.\ (\ref{factor2}) acts as a suppression factor.
Specifically, the eigenvalue with the biggest real 
part will drive the 
strongest suppression for the component of the scattering
oriented along the direction of the corresponding eigenvector.
As we shall see, most of the relevant information about 
the $\Delta \eta$ behavior of the cross section 
is derived from the properties
of the soft anomalous dimension matrix 
$\Gamma^{(\a)}_S(\alpha_s)$, which we will study 
in detail in Sec.\ \ref{anodim}.

\section{The Gap Fraction}
\label{fraction}
We will be interested in the evaluation of the gap fraction, 
defined as the ratio of the number of dijet events with a specified rapidity gap
to the total number of dijet events.
A gap event is usually identified experimentally by the lack of particle multiplicity
in the interjet region \cite{D0CDF,ZEUS}.
\begin{figure}[t]
\begin{center}
\mbox{\epsfysize=8.6cm \epsfxsize=8.6cm \epsfbox{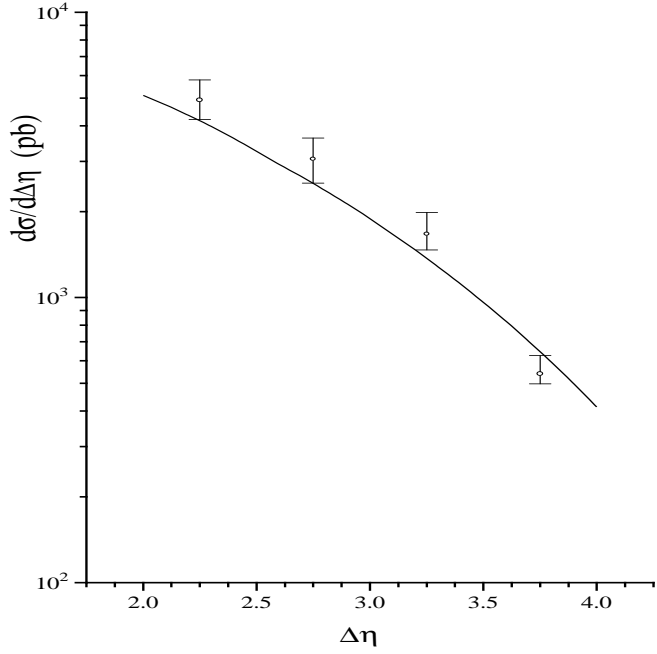}}
\vspace{.3cm}
\caption[dum]{The overall dijet  cross section compared with the experimental data of Ref.\ \cite{ZEUS}.\label{figLOovall}}
\end{center}
\end{figure}
The multiplicity is determined from the number of  
calorimeter cells which measure energy deposition above a threshold,
$300$ MeV in the case of Ref.\ \cite{ZEUS}.
Therefore, the condition for a gap event is the absence
of such cells (or clusters of cells) in the rapidity region between the jets.  
Our formulation of the problem, Eq.\ (\ref{factcs1}), 
is in terms of the total interjet flow, $Q_c$, of hadronic radiation, which is 
clearly related to particle multiplicity. 
In analogy with experiment, we introduce an energy  threshold, $Q_0$, 
which is different in principle from the experimental calorimeter threshold,
and identify a gap event from the condition of interjet radiation
less than $Q_0$.
From Eqs. (\ref{factcs1}), (\ref{crosdir}), (\ref{crosres}) we see 
that the dependence on the interjet radiation is all in the partonic scattering, 
Eq.\ (\ref{factor2}).
\begin{figure}[t]
\begin{center}
\mbox{\epsfysize=8.6cm \epsfxsize=8.6cm \epsfbox{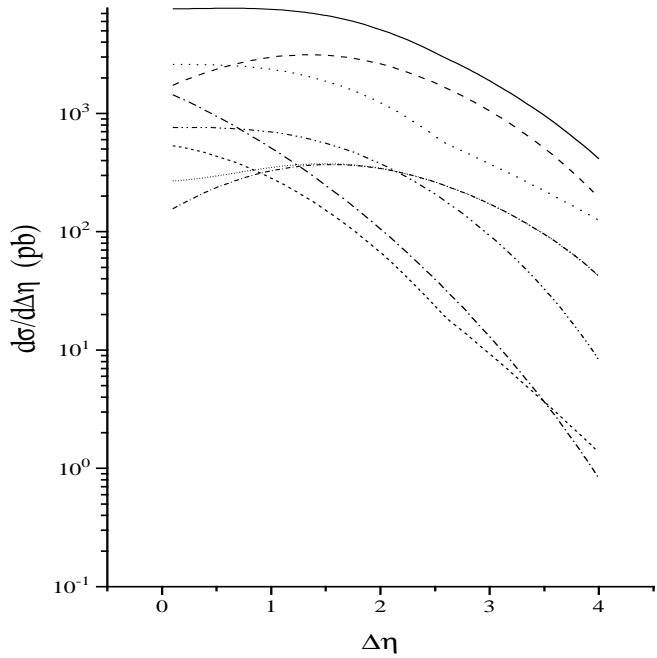}}
\vspace{.3cm}
\caption[dum]{The contribution of the different partonic reactions 
to the dijet cross section. At $\Delta \eta = 0$ we can identify
from top to bottom: 
the overall result (solid line); the contributions of:
$\gamma +g \rightarrow q +\bar{q}$ (dotted line) ,
$q+g \rightarrow q+g$  (dashed line),
$q+g \rightarrow g+q$  (dot dashed line),
$g+g \rightarrow g+g$ (double dot dashed line),
$\gamma +q(\bar{q}) \rightarrow g +q(\bar{q})$ (short dashed line),
$q+q \rightarrow q+q$  (short dotted line),
$q+\bar{q} \rightarrow q+\bar{q}$  (short dot dashed line).
Here the two reactions $q+g \rightarrow q+g$
and $q+g \rightarrow g+q$ differ from each other by the exchange
of the Mandelstam invariants $\hat{t}$ and $\hat{u}$ (see Eqs.\ 
(\ref{qgqganalcsl}) and (\ref{qggqanalcsl}) in the Appendix).
\label{figLOpart}}
\end{center}
\end{figure}
Then, in order to get a dijet gap cross section,
we just need to evaluate Eq.\ (\ref{factor2}) at the threshold value $Q_0$  
of the interjet energy flow  $Q_c$,
\beqa
\frac{d\hat{\sigma}_{\gamma p}
^{(\a), {\rm{gap}}}}{d \Delta \eta }\left(Q_0, \hat{s},\hat{t},\eta_{JJ},\Delta y,\alpha_s(\hat t)\right)
&=&\, \frac{\pi}{2\hat{s}} \, \left( 2 \cosh^2 \left( \frac{\Delta \eta}{2} \right)\right)^{-1} 
\nonumber \\
&& \hspace{-40mm} \times \sum_{\alpha, \beta} H^{(\a,1)}_{\beta \alpha}\left( \Delta y,\sqrt{\hat{s}},\sqrt{-\hat{t}},
\alpha_s\left(\hat{t}\right) \right) 
 S^{{(\a,0)}}_{\alpha \beta} ( \Delta y ) \,  \nonumber \\
&& \hspace{-40mm} \times \, 
\left[\ln\left({Q_0\over \Lambda}\right)\right]^{E^{(\a)}_{\alpha\beta}}\;  
\left[ \ln \left( {\sqrt{-\hat{t}}\over \Lambda}\right)
\right]^{-E^{(\a)}_{\alpha\beta}} . \nonumber\\
\label{nume}
\eeqa
The denominator of the fraction is also obtained 
from Eq.\ (\ref{factor2}), this time evaluated at the maximum value of $Q_c$, 
the hard scale $\sqrt{-\hat{t}}$.
Values of $Q_c$ above this scale would break the premise
underlying the factorization in Eq.\ (\ref{factor}), since
the emission into the interjet region would not be soft any more.
By performing the substitution $Q_c=\sqrt{-\hat{t}}$ in Eq.\ (\ref{factor2})
we find the full leading order partonic dijet cross section
\beqa 
\frac{d\hat{\sigma}_{\gamma p}^{(\a)}}{d\Delta\eta}
\left(\hat{s},\hat{t},\eta_{JJ},\Delta y,\alpha_s(\hat t)\right)
&=&\, \frac{\pi}{2\hat{s}} \, \left( 2 \cosh^2 \left( \frac{\Delta \eta}{2} \right)\right)^{-1}
\nonumber \\
&& \hspace{-40mm} \times H^{(\a,1)}_{\beta \alpha}
\left( \Delta y,\sqrt{\hat{s}},\sqrt{-\hat{t}},
\alpha_s\left(\hat{t}\right) \right) 
 S^{{(\a,0)}}_{\alpha \beta} ( \Delta y ) \,, 
\label{deno}
\eeqa
where, from Eqs. (\ref{eq:newbasS}) and (\ref{newbasH}), 
for each partonic process the trace identity
$H^{(\a,1)}_{\beta \alpha} \, S^{{(\a,0)}}_{\alpha \beta}=
H^{(\a,1)}_{IL} \, S^{{(\a,0)}}_{LI}$ holds,
which is proportional to the leading order partonic cross section (see also 
Sec.\ \ref{hardsoft} and
Appendix \ref{AppA}).
After summing the gap and total cross sections
over partonic subprocesses, as in Eqs.\ (\ref{crosdir}) and (\ref{crosres}), 
the gap fraction has the form
\beq
f^{{\rm{gap}}}={{d\sigma^{{\rm{gap}}}_{ep}\over
d\Delta\eta}(Q_0,S,E_T,\Delta y) \over
{d\sigma_{ep}\over d\Delta\eta}(S,E_T,\Delta y)}\,.
\label{fr}
\eeq
In the numerical evaluation of the denominator of this fraction,
we use, for each partonic reaction $\a$ in Eq.\ (\ref{deno}), 
the invariant S-matrix elements squared, and corresponding  
parton luminosities, $L^{(\a)}$, summarized for
completeness in Appendix \ref{AppA}.
We have used four quark flavors, $n_f=4$, and 
have assumed flavor symmetry for the sea quarks.
The numerical result we have found, shown in Fig.\ \ref{figLOovall}, is in
reasonable agreement with the data points from Ref.\ \cite{ZEUS}.
Here it should be emphasized that, for the overall
dijet cross section, and also, as we will see in Sec.\ \ref{results}, for the 
gap cross section, corrections are expected
from next-to-leading order contributions to the hard matrix \cite{photold,Aur,FRGS,KK}. 
However, the gap fraction should not be too sensitive to them.

\begin{figure}[t]
\begin{center}
\mbox{\epsfysize=8.6cm \epsfxsize=8.6cm \epsfbox{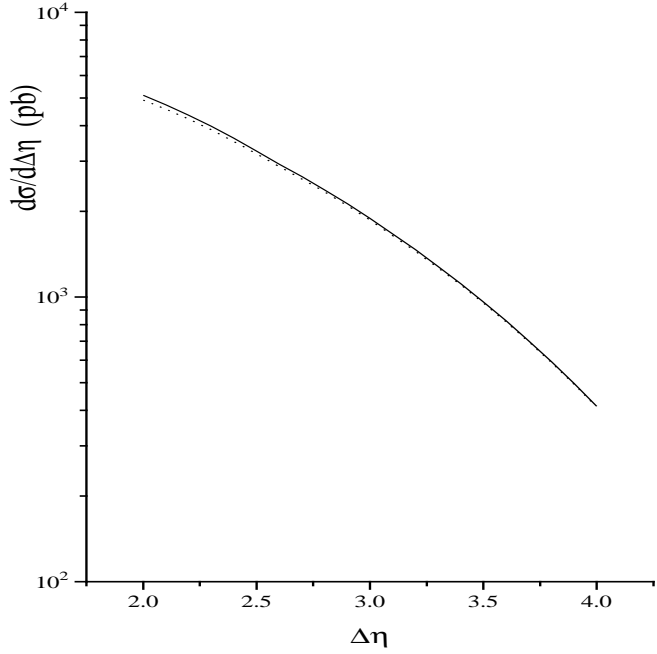}}
\vspace{.3cm}
\caption[dum]{The dijet cross section (solid line) and the result obtained by 
dropping from Fig.\ \ref{figLOpart} the contributions of the direct process
$\gamma +q(\bar{q}) \rightarrow g +q(\bar{q})$, 
and  of the resolved reaction $q+g \rightarrow g+q$ (dotted line).
\label{figLO4proc}}
\end{center}
\end{figure}

In Fig.\ \ref{figLOpart} we show the partonic reactions
most relevant to the overall cross section, evaluated on
the full $\Delta \eta$-range. 
For the $\Delta \eta$-range
of Ref.\ \cite{ZEUS}, $\Delta \eta >2.0$, Fig.\ \ref{figLO4proc} clearly
shows that it is sufficient to
consider the contributions from 
the direct process, $\gamma + g \rightarrow q + \bar{q}$, the
resolved gluon-initiated 
reactions, $q+g \rightarrow q+g$ and $g+g \rightarrow g+g$,
and the quark processes, 
$q+\bar{q} \rightarrow q+\bar{q}$ and $q+q \rightarrow q+q$.
These will also be the dominant contributions to 
the numerator of the fraction, which
differs from the denominator only by the last
two factors in Eq.\ (\ref{nume}), depending 
on the soft dynamics of each specific partonic reaction. 
The determination of the gap cross section will require 
an analysis of the partonic scattering in terms of color flow, 
to be found in the next
Section, and the knowledge of the spectrum of the process-dependent 
soft anomalous 
dimension matrices, to be discussed in Sec.\ \ref{anodim}.

\section{Lowest Order Hard and Soft Matrices in Color Space}
\label{hardsoft}
In this section, we present the color decomposition of the hard scattering
for the resolved partonic processes that have been shown above  
to give the dominant contribution to the dijet cross section in the $\Delta \eta$
region of interest.
For each case, we will define a basis of color tensors, $\left\{ |{c^{(\a)}}_I \rangle \right\}$,
and give, in this basis, the Born level hard matrix, $H^{(\a,1)}_{IL}$, whose
elements come from the squares of the color-decomposed tree amplitudes.
We will also give the zeroth order soft matrix, $S^{{(\a,0)}}_{LI}$, defined 
by the set of traces $S^{{(\a,0)}}_{LI}=\rm{Tr}\left[ \left( c^{{(\a)}}_{L}\right)^{\dagger} \, c^{{(\a)}}_{I} \right]$.
The knowledge of these matrices is not only crucial for the evaluation of Eqs.\ (\ref{nume}) and (\ref{deno}),
but will be important for threshold resummation in jet production as well \cite{KLOS}.

For the color tensors, following Ref.\ \cite{KOS2}, we will find it convenient to use $t$-channel bases.
In terms of momentum and color, we label the partonic process
$\a$ according to $f_{\gamma}\left( l_A, \e_A \right)
+f_p \left( l_B, \e_B \right) 
\rightarrow f_1\left( p_1, \e_1 \right)+
f_2\left( p_2, \e_2 \right)$, 
where $\{l_A, \e_A\}$, $\{l_B, \e_B\}$ ($\{p_1, \e_1\}$, $\{p_2, \e_2\}$) are, respectively,
the 
momenta and colors of the incoming (outgoing)
partons. 
The hard matrix will be expressed in terms of the partonic Mandelstam invariants
\beqa
\hat{s}&=&\left( l_A+l_B \right)^2 \nonumber \\
\hat{t}&=&\left( l_A-p_1 \right)^2 \nonumber \\
\hat{u}&=&\left( l_A-p_2 \right)^2 \, .
\label{Mandlst}
\eeqa
It will also depend on
the QCD running coupling evaluated at the hard scale, $(-\hat{t})^{1/2}$,
according to the formula
\beq 
\alpha_s\left(\hat{t}\right)=\frac{2\pi}{\beta_1 \ln \left( (-\hat{t})^{1/2}/ \Lambda \right)} \, .
\label{QCDcoupl}
\eeq
For each process, it is  straightforward to check that the color trace over the product of hard and 
soft matrices gives, as expected, the tree level invariant matrix element squared 
of Appendix \ref{AppA} \cite{ESW}.

\subsection{Hard and soft matrix for $qg \rightarrow qg$}
For the partonic process  $q+g \rightarrow q+g$ we define the basis of color tensors \cite{KOS2}
\beqa
c_1&=&\delta_{\e_A,\e_1}\delta_{\e_B,\e_2}\nonumber\\
c_2&=&d_{\e_B \e_2 c}{\left( T_F^c \right)}_{\e_1 \e_A}\nonumber\\
c_3&=&if_{\e_B \e_2 c}{\left( T_F^c \right)}_{\e_1 \e_A} \, ,
\label{qgqgbasis}
\eeqa 
where $c_1$ is the $t$-channel singlet tensor, $c_2$ and $c_3$ the
symmetric and antisymmetric octet respectively.
Here and below, we suppress color indices on the $c_I$'s.
In this basis, the Born level hard scattering can be described in color space by the matrix 
\beq
H^{(1)}\left(\sqrt{-\hat{t}},
\sqrt{\hat{s}},\alpha_s(\hat{t}) \right)=\frac{1}{48} \, \alpha^2_s(\hat{t}) 
\left( 
\begin{array}{ccc}
\frac{1}{9}\chi_1    &
\frac{1}{3}\chi_1   &
\frac{2}{3}\chi_2  \vspace{2mm} \\
 \frac{1}{3}\chi_1  & \chi_1 & 2\chi_2 \vspace{2mm} \\
 \frac{2}{3} \chi_2 & 2\chi_2 & \chi_3 
\end{array}
\right) \, ,
\eeq
with $\chi_1$, $\chi_2$ and $\chi_3$ functions of the partonic
Mandelstam invariants:
\beqa
\chi_1&=&2- \frac{\hat{t}^2}{\hat{s}\hat{u}}      \nonumber \\
\chi_2&=&1-\frac{\hat{s}}{\hat{t}}-\frac{\hat{u}^2}{\hat{s}\hat{t}} - 
\frac{1}{2}\frac{\hat{t}^2}{\hat{s}\hat{u}} 	\nonumber \\
\chi_3&=&6-8\frac{\hat{s} \hat{u}}{\hat{t}^2}-\frac{\hat{t}^2}{\hat{s}\hat{u}} \,.
\eeqa
It is easy to see that, in the limit of forward scattering ($\Delta \eta \rightarrow \infty$),
the component of the hard matrix in the color direction of  the antisymmetric octet
becomes dominant, since
it is ${\cal{O}}(\hat{t}^{-2})$.

The zeroth order soft matrix, on the other hand,  is given by
\beq
S^{(0)}=\left( 
\begin{array}{ccc}
N_c(N_c^2-1)  & 0 & 0 \\
0 & \frac{1}{2N_c}(N_c^2-4)(N_c^2-1) &0\\
0 & 0 &\frac{1}{2} N_c(N_c^2-1)  
\end{array}
\right) \, ,
\eeq
where $N_c=3$ is the number of colors.

\subsection{Hard and soft matrix for $gg \rightarrow gg$}
For this process, a suitable $t-$channel basis of color 
tensors has been defined in Ref.\ \cite{KOS2}
to be:
\beqa
c_1&=&\frac{i}{4}\left[f_{\e_A \e_B l}
d_{\e_1 \e_2 l}-d_{\e_A \e_B l}f_{\e_1 \e_2 l}\right]\, ,
\nonumber \\
c_2&=&\frac{i}{4}\left[f_{\e_A \e_B l}
d_{\e_1 \e_2 l}+d_{\e_A \e_B l}f_{\e_1 \e_2 l}\right] \, ,
\nonumber \\
c_3&=&\frac{i}{4}
\left[f_{\e_A \e_1 l}d_{\e_B \e_2 l}+d_{\e_A \e_1 l}f_{\e_B \e_2 l}\right] \, ,
\nonumber \\
c_4&=&P_1(\e_A,\e_B;\e_1,\e_2)=\frac{1}{8}\delta_{\e_A \e_1} \delta_{\e_B \e_2} \, ,
\nonumber \\ 
c_5&=&P_{8_S}(\e_A,\e_B;\e_1,\e_2)=\frac{3}{5} d_{\e_A\e_1c} d_{\e_B\e_2c} \, ,
\nonumber \\ 
c_6&=&P_{8_A}(\e_A,\e_B;\e_1,\e_2)=\frac{1}{3} f_{\e_A\e_1c} f_{\e_B\e_2c} \, ,
\nonumber \\ 
c_7&=&P_{10+{\overline {10}}}(\e_A,\e_B;\e_1,\e_2)=
\frac{1}{2}(\delta_{\e_A \e_B} \delta_{\e_1 \e_2}
-\delta_{\e_A \e_2} \delta_{\e_B \e_1}) \nonumber \\
&& \hspace{5.1cm}-\frac{1}{3} f_{\e_A\e_1c} f_{\e_B\e_2c} \, ,
\nonumber \\ 
c_8&=&P_{27}(\e_A,\e_B;\e_1,\e_2)=
\frac{1}{2}(\delta_{\e_A \e_B} \delta_{\e_1 \e_2}
+\delta_{\e_A \e_2} \delta_{\e_B \e_1})
\nonumber \\
&& \hspace{4.5cm}-\frac{1}{8}\delta_{\e_A \e_1} \delta_{\e_B \e_2}
-\frac{3}{5} d_{\e_A\e_1c} d_{\e_B\e_2c} \, .
\label{ggggbasis}
\eeqa 
The last five elements of the basis are the $t$-channel $SU(3)$
projectors for the decomposition into irreducible representations
of the direct product $8 \otimes 8$, which corresponds to the color content of a set of two gluons.

In this basis, the hard matrix has the block-diagonal structure
\beqa
 H^{(1)}\left(\sqrt{-\hat{t}},
\sqrt{\hat{s}},\alpha_s(\hat{t}) \right)=\left(\begin{array}{cc}
	     0_{3 \times 3}        & 0_{3 \times 5} \\
	      0_{5 \times 3}      &  H^{(1)}_{5 \times 5}
\end{array} \right)\, ,
\label{hardblock}
\eeqa
where the $5 \times 5$ submatrix $H^{(1)}_{5 \times 5}$ is given by
\beqa
H^{(1)}_{5 \times 5}\left(\sqrt{-\hat{t}},
\sqrt{\hat{s}},\alpha_s(\hat{t}) \right)=\frac{1}{16} \, \alpha^2_s(\hat{t})
\left(\begin{array}{ccccc}
9\chi_1 & \frac{9}{2}\chi_1 & \frac{9}{2}\chi_2 & 0 & -3\chi_1 \vspace{2mm}\\
\frac{9}{2}\chi_1 & \frac{9}{4}\chi_1 & \frac{9}{4}\chi_2 & 0 & -\frac{3}{2}\chi_1 \vspace{2mm}\\
\frac{9}{2}\chi_2 & \frac{9}{4}\chi_2 & \chi_3 & 0 & -\frac{3}{2}\chi_2 \vspace{2mm}\\
0 & 0 & 0 & 0 & 0 \vspace{2mm}\\
-3\chi_1 & -\frac{3}{2}\chi_1 & -\frac{3}{2}\chi_2 &0 &\chi_1 
\end{array}\right) \, ,
\eeqa
with $\chi_1$, $\chi_2$ and $\chi_3$  defined by
\beqa
\chi_1&=&1-\frac{\hat{t}\hat{u}}{\hat{s}^2}-\frac{\hat{s}\hat{t}}{\hat{u}^2}+
\frac{\hat{t}^2}{\hat{s}\hat{u}}      \nonumber \\
\chi_2&=&\frac{\hat{s}\hat{t}}{\hat{u}^2}-\frac{\hat{t}\hat{u}}{\hat{s}^2}+
\frac{\hat{u}^2}{\hat{s}\hat{t}}-
\frac{\hat{s}^2}{\hat{t}\hat{u}} 	\nonumber \\
\chi_3&=&\frac{27}{4}-9\left(\frac{\hat{s}\hat{u}}{\hat{t}^2}+\frac{1}{4}
\frac{\hat{t}\hat{u}}{\hat{s}^2}+\frac{1}{4}\frac{\hat{s}\hat{t}}{\hat{u}^2}\right)
+\frac{9}{2}\left(\frac{\hat{u}^2}{\hat{s}\hat{t}}+\frac{\hat{s}^2}{\hat{t}\hat{u}}-
\frac{1}{2}\frac{\hat{t}^2}{\hat{s}\hat{u}}\right) \,.
\eeqa
Once again, it can be noticed that the component of the hard scattering 
in the color direction corresponding to the antisymmetric octet dominates
in the forward limit.

For the zeroth order soft matrix, straightforward color traces with $N_c=3$ colors give
\beq
S^{(0)}=\left( 
\begin{array}{cccccccc}
-5 & 0  & 0 & 0 & 0 & 0 & 0 & 0   \\
0  & -5 &0 & 0 & 0 &0 & 0 & 0    \\
0  & 0  &-5 & 0 & 0 &0 & 0 & 0     \\
0  & 0  & 0 & 1 & 0 &0 & 0 & 0      \\  
0  & 0  & 0 & 0 & 8 &0 & 0 & 0      \\
0  & 0  & 0 & 0 & 0 &8 &0 &0      \\
0  & 0  & 0 & 0 & 0 &0 &20 &0     \\
0  & 0  & 0 & 0 & 0 &0 &0 &27     
\end{array}
\right) \, .
\label{ggggsoft}
\eeq
We notice that the first three color tensors of the basis, Eq.\ (\ref{ggggbasis}), have negative eigenvalues,
while the eigenvalues of the projectors count the number of color states
belonging to each  irreducible representation. 
In fact, $c_1$, $c_2$ and $c_3$ of Eq.\ (\ref{ggggbasis}) decouple from the physical cross section.
Eq.\ (\ref{hardblock}) supports this interpretation, since
the components of the Born level hard scattering along these
color directions vanish identically, 
reducing the dimensionality of the problem, in color space, 
from $8 \times 8$ to $5 \times 5$.
In principle, from Eqs.\ (\ref{eq:newbasS}) and (\ref{newbasH}), 
the original dimensionality could be restored
after the change from the color basis to the basis of the eigenvectors
of the  soft anomalous dimension matrix, $\Gamma_S$. 
We will show below that the structure of  the specific $\Gamma_S$ for $g+ g \rightarrow g +g$
prevents this from happening.

\subsection{Hard and soft matrix for $q\bar{q} \rightarrow q\bar{q}$ and $qq \rightarrow qq$}
Both processes can be conveniently treated in
the $t-$channel singlet-octet basis \cite{KOS2}
\beqa
 {c_1}&=&\delta_{\e_A,\e_1}\delta_{\e_B,\e_2}\nonumber\\
 {c_2}&=&-\frac{1}{2N_c}\delta_{\e_A,\e_1}
\delta_{\e_B,\e_2}+\frac{1}{2}\delta_{\e_A,\e_B}\delta_{\e_1,\e_2}.
\label{qqbbasis}
\eeqa 

In color space, the hard matrix has the form
\beq
H^{(1)}\left(\sqrt{-\hat{t}},
\sqrt{\hat{s}},\alpha_s(\hat{t}) \right)=\frac{1}{9} \, \alpha^2_s(\hat{t}) \left( 
\begin{array}{cc}
\frac{32}{81} \chi_1  & \frac{1}{9} \chi_2 \vspace{2mm} \\
\frac{1}{9} \chi_2 & \chi_3
\end{array}
\right) \,,
\eeq
where, in the case of $q\bar{q} \rightarrow q\bar{q}$, $\chi_1$, $\chi_2$ and $\chi_3$  are
defined by
\beqa
\chi_1&=&\frac{\hat{t}^2+\hat{u}^2}{\hat{s}^2}      \nonumber \\
\chi_2&=&8\frac{\hat{u}^2}{\hat{s}\hat{t}}-
\frac{8}{3}\frac{\hat{t}^2+\hat{u}^2}{\hat{s}^2} \nonumber \\
\chi_3&=&2\frac{\hat{s}^2+\hat{u}^2}{\hat{t}^2}+
\frac{2}{9}\frac{\hat{t}^2+\hat{u}^2}{\hat{s}^2}-\frac{4}{3}\frac{\hat{u}^2}{\hat{s}\hat{t}}\,,
\eeqa
while in the case of $qq \rightarrow qq$, related to $q\bar{q} \rightarrow q\bar{q}$ by 
the crossing transformation $\hat{s} \leftrightarrow \hat{u}$, they are given by
\beqa
\chi_1&=&\frac{\hat{t}^2+\hat{s}^2}{\hat{u}^2}      \nonumber \\
\chi_2&=&8\frac{\hat{s}^2}{\hat{t}\hat{u}}-
\frac{8}{3}\frac{\hat{s}^2+\hat{t}^2}{\hat{u}^2}  \nonumber \\
\chi_3&=&2\frac{\hat{s}^2+\hat{u}^2}{\hat{t}^2}+
\frac{2}{9}\frac{\hat{s}^2+\hat{t}^2}{\hat{u}^2}-\frac{4}{3}\frac{\hat{s}^2}{\hat{t}\hat{u}}\,.
\eeqa
In both cases, for forward scattering, the component of the hard matrix along the color octet direction
becomes dominant. 
\begin{figure}[t]
\begin{center}
\mbox{\epsfysize=10.cm \epsfxsize=8.6cm \epsfbox{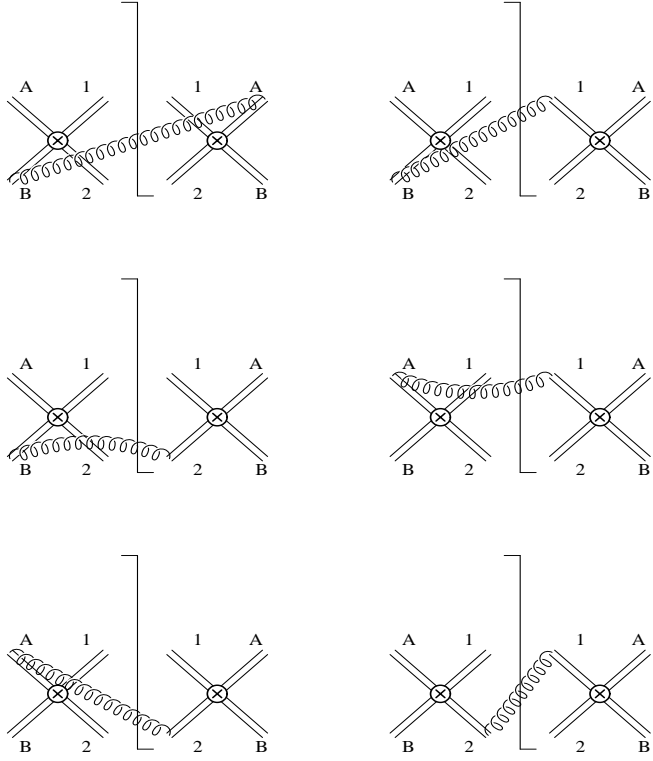}}
\vspace{.3cm}
\caption[dum]{Real corrections to the eikonal 
scattering, Eq.\ (\ref{eq:eikcs}). For brevity, 
we only show half of the contributing cut diagrams. The 
remaining ones can be obtained by hermitian conjugation of  each of the above 
graphs, i.\ e.\ , by reflection with respect to the final state cut.
\label{eikdiag1}}
\end{center}
\end{figure}

For the lowest order soft matrix, after the color trace, 
we find the common simple form \cite{StOd1} 
\beq
S^{(0)}=\left( 
\begin{array}{cc}
N_c^2  & 0  \\
0 &  \frac{1}{4}(N_c^2-1)  
\end{array}
\right) \, .
\eeq

\section{The Soft Anomalous Dimension Matrix}
\label{anodim}
In this section we shall describe how to compute     
the process-dependent soft anomalous dimension matrix, $\Gamma^{(\a)}_S$ \cite{StOd}.
The $\mu$-independence of the left hand side of Eq.\ (\ref{factor})
implies that the matrices $H^{(\a)}_{IL}$ and  $S^{(\a)}_{LI}$ renormalize multiplicatively,
\beqa
H^{(\a)}{}^{(B)}_{IL}&=& \left(Z_S^{(\a)}{}^{-1}\right)_{IC}H_{CD}^{(\a)}\; 
[(Z_S^{(\a)}{}^\dagger)^{-1}]_{DL}\cr
S^{(\a)}{}^{(B)}_{LI}&=&(Z_S^{(\a)}{}^\dagger)_{LB}S_{BA}^{(\a)}Z_{S,AI}^{(\a)}{}\, ,
\label{eq:barereno}
\eeqa
where the superscript $(B)$ identifies the bare quantities, and  $Z^{(\a)}_{S,CD}$ is a matrix of
renormalization constants,  describing
the renormalization of the soft function, i.\ e.\ , the eikonal scattering, Eq.\ (\ref{relSeikcs}).
The one-loop anomalous dimension $\Gamma^{(\a)}_S$ is obtained from the residue of the UV pole
contained in the matrix $Z^{(\a)}_S$,
\beq
\left(\Gamma^{(\a)}_S\right)_{LI}(g)=-\frac{g}{2} \frac{\partial}
{\partial g} {\rm{Res}}_{\epsilon \rightarrow 0} 
\left(Z^{(\a)}_S\right)_{LI}(g,\epsilon).
\label{gammadef}
\eeq
From Eq.\  (\ref{eq:eikcs}) we see that 
the potential sources of UV divergences are the virtual vertex corrections
to the eikonal color-dependent operators $w_I(x)_{\{c_k\}}$, and also the real corrections, 
when gluons are emitted 
into the forward region of the scattering,
because the eikonal cross section, Eq.\ (\ref{eq:eikcs}), is completely
inclusive with respect to the amount of forward radiation.
\begin{figure}[t]
\begin{center}
\mbox{\epsfysize=10.cm \epsfxsize=8.6cm \epsfbox{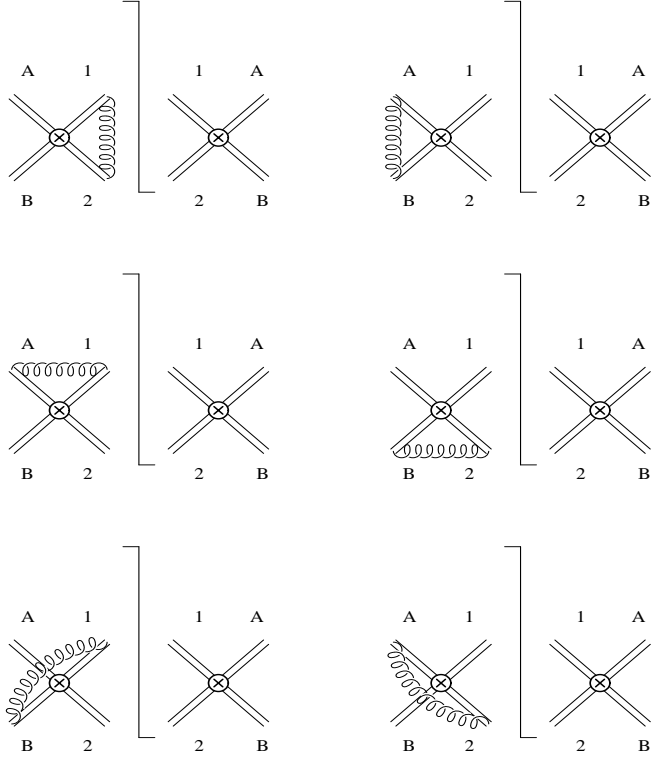}}
\vspace{.3cm}
\caption[dum]{Virtual corrections to the eikonal scattering, Eq.\ (\ref{eq:eikcs}). For brevity, 
we only show half of the contributing cut diagrams. The 
remaining ones can be obtained by hermitian conjugation of  each of the above 
graphs, i.\ e.\ , by reflection with respect to the final state cut.\label{eikdiag2}}
\end{center}
\end{figure}

The Feynman rules for the evaluation of eikonal diagrams have been presented in
Ref.\ \cite{KOS2}. 
The necessary one-loop calculations are illustrated
in Figs.\ \ref{eikdiag1}-\ref{eikdiag2}:
given the two color vertices $c_I$ and $c_L$,
corresponding to the hard amplitude 
and to its complex conjugate, we build the eikonal cross section from the product
of the color-dependent eikonal operators, $w_I(x)_{\{c_k\}}$ and 
$\left(w_L(x)_{\{c_k\}}\right)^{\dagger}$,
Eq.\ (\ref{eq:eikcs}), and
we compute the diagrams  obtained by adding one extra real or
virtual gluon, shown respectively in Fig.\ \ref{eikdiag1} and \ref{eikdiag2}. 
Working
in Feynman gauge, the self-energy diagrams do not contribute, because
they are proportional to the invariant mass 
of each lightlike eikonal line.

From the loop integration, it is easy to see
that the UV divergences from the gluon emission diagrams and 
from the virtual diagrams with the gluon emitted and reabsorbed between an incoming and 
an outgoing eikonal line, are real.
On the other hand,  the coefficient of the UV pole from 
virtual diagrams 
with the gluon emitted and reabsorbed between two incoming or two outgoing
eikonal lines, also have imaginary parts.
Since the imaginary part of  the soft anomalous dimension matrix comes only from
virtual eikonal diagrams, it can be extracted directly from
Ref.\ \cite{KOS2}, where contributions to $\Gamma_S$ are computed, for most partonic processes, 
from the virtual corrections
to the eikonal operators $w_I(x)_{\{c_k\}}$.
The real parts of the UV divergences coming, 
respectively, from virtual and real emission diagrams
partially cancel in the sum, leaving a remainder which is a function of 
$\Delta y$, the width of the
interjet rapidity region, where the gluon radiation
is measured, and of the scattering-dependent parameter $\Delta \eta$. 

We will now present, for each of the partonic processes relevant to the dijet cross
section, the result for   $\Gamma^{(\a)}_S$, and the analysis 
of its eigenvalues and eigenvectors, which we will relate 
to the behavior of the cross section, Eq.\ (\ref{factor2}).
For the direct reaction $\gamma + g \rightarrow q +\bar{q}$, which has
only color octet content, $\Gamma^{(\a)}_S$ will 
reduce to a function, related by crossing to the soft anomalous dimension
computed in Ref.\ \cite{LOS} for direct photon production. 

We will see below, for all the resolved reactions, that the eigenvectors of
$\Gamma_S$ are mixed color states, and that their color composition
is a function of the gap width only, through the ``geometrical'' parameter $\Delta y$. 
We have already pointed out in Eq.\ (\ref{dydeta}) that, in the experimental
configuration of Ref.\ \cite{ZEUS}, $\Delta \eta$ fixes $\Delta y$,
leaving us with only one independent parameter. 
We should emphasize, however, that for the case of the Tevatron rapidity gaps
\cite{StOd1,StOd}, where the gap is taken fixed in the calorimeter
detector, and is independent of the dynamics of the scattering, the eigenvectors have 
the same form as for this problem. 
Below, we will find it convenient to refer to the eigenvectors as ``quasi-color'' states
(quasi-octet, quasi-singlet, and so on), 
identified by their behavior
in the limit of  asymptotically large gap regions, where they reduce 
to pure color states.
We will discover that the eigenvalue with the smallest real part always 
corresponds to the quasi-singlet eigenvector, and we will show,
from the discussion in Sec.\ \ref{refacto}, that the related
component of the hard scattering is the only one to survive for large
values of $\Delta y$.

\subsection{Soft anomalous dimension for  $qg \rightarrow qg$}
In the basis of Eq.\ (\ref{qgqgbasis}),
the one-loop soft anomalous dimension matrix is
\beqa
{\Gamma_S}\left(\Delta \eta,
\Delta y \right)=\frac{\alpha_s}{2\pi} \left(
\begin{array}{ccc}
\rho_1+\xi_1 & 0 & 2i\pi \\
0 & \rho_1 & N_c i \pi \\
4i\pi & \frac{N_c^2-4}{N_c}i\pi & \rho_1 
\end{array} \right),
\label{qgqgano}
\eeqa
where the functions $\rho_1$ and $\xi_1$ are defined as follows:
\beqa 
&&\hspace{-1cm} \g_1( {\Delta y})=N_c
\left(i\pi-2 {\Delta y}\right),  \nonumber \\ 
&&\hspace{-1cm} \w_1(\Delta \eta,
 {\Delta y})=N_c \left[2 \ln \left( \frac{\tanh(\frac{\Delta \eta}{2})+
\tanh(\frac{\Delta y}{2})}
{\tanh(\frac{\Delta \eta}{2})-\tanh\left( 
\frac{ {\Delta
 y}}{2}\right)}\right)\right] +i\pi \frac{2N_c^2-1}{N_c} \nonumber \\
&&\hspace{-1cm}-\frac{N_c^2+1}{2N_c}\left[\ln
\left(\frac{\tanh(\frac{\Delta \eta}{2})+\tanh\left( \frac{
 {\Delta y}}{2} 
\right)}
{\tanh(\frac{\Delta \eta}{2})
-\tanh\left( \frac{ {\Delta y}}{2} 
\right)}\right) + \ln
\left(\frac{1-\tanh\left(\frac{ {\Delta y}}{2}
\right)}
{1
+\tanh\left(\frac{ {\Delta y}}{2}
\right)}\right)\right]\, .
\label{xi1rho1}
\eeqa
While the first function depends only on the rapidity width of the
interjet region, the second also depends on the hard scattering through 
$\Delta \eta$.

The eigenvalues of the above matrix are given by 
\beqa
\lambda_1(\Delta \eta,
 {\Delta y})&=&\frac{\alpha_s}{2\pi} \left[
\frac{2  {\rho_1}+ {\xi_1}}{3}
+e^{\frac{2i\pi}{3}}\left( {s_1}+ {s_2}\right)
\right]\nonumber \\
\lambda_2(\Delta \eta,
 {\Delta y})&=&\frac{\alpha_s}{2\pi} \left[
\frac{2  {\rho_1}+ {\xi_1}}{3}
+\left(e^{\frac{4i\pi}{3}} {s_1}+ {s_2}\right)
\right]\nonumber \\
\lambda_3(\Delta \eta,
 {\Delta y})&=&\frac{\alpha_s}{2\pi} \left[
\frac{2  {\rho_1}+ {\xi_1}}{3}
+\left( {s_1}+e^{\frac{4i\pi}{3}} {s_2}\right)
\right] \, ,
\label{qgqgeigval}
\eeqa
where $s_1$ and $s_2$, in the $\Delta y$ range of Ref.\ \cite{ZEUS}
\footnote{The definitions of $s_1$ and $s_2$ given in Eqs.\ (\ref{s1}) and
(\ref{s2}) are valid for $\Delta y< 4\pi \frac{\sqrt{10}}{15}$, 
the branch point
of the square root on the right hand side of Eqs.\ (\ref{s1}) and
(\ref{s2}) . For larger
values of $\Delta y$, the square root picks
the second branch definition. Consequently, $s_1$ and $s_2$ are switched.}, $\Delta y < 2$,
are the functions
\beqa
s_1(\Delta y)&\equiv&\left[\frac{ {\xi_1}^3}{27}+\frac{\pi^2}{3}(N_c^2-8) {\xi_1}+
\left(\frac{\pi^2}{27}(N_c^2-4) {\xi_1}^4+ \right. \right. \nonumber \\
&&\left. \left. \frac{2\pi^4}{27} {\xi_1}^2\left(
N_c^4-28N_c^2+88\right)+\frac{\pi^6}{27}(N_c^2+4)^3\right)^{\frac{1}{2}}
\right]^{\frac{1}{3}} \, ,
\label{s1}
\eeqa
and
\beqa
s_2(\Delta y)&\equiv& \left[\frac{ {\xi_1}^3}{27}+\frac{\pi^2}{3}(N_c^2-8) {\xi_1}-
\left(\frac{\pi^2}{27}(N_c^2-4) {\xi_1}^4+ \right. \right. \nonumber \\
&&\left. \left. \frac{2\pi^4}{27} {\xi_1}^2\left(
N_c^4-28N_c^2+88\right)+\frac{\pi^6}{27}(N_c^2+4)^3\right)^{\frac{1}{2}}
\right]^{\frac{1}{3}} \, . 
\label{s2}
\eeqa
\begin{figure}[t]
\begin{center}
\mbox{\epsfysize=6cm \epsfxsize=15cm \epsfbox{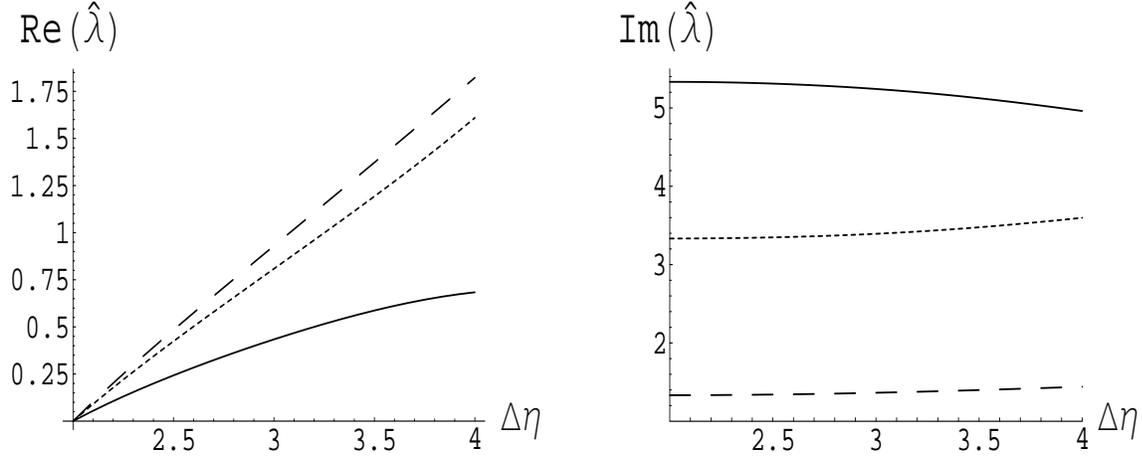}}
\vspace{.3cm}
\caption[dum]{Plot of the real (left) and imaginary (right) part of the eigenvalues ($\hat{\lambda}_i=\lambda_i/\alpha_s$)
of the soft anomalous dimension matrix for $q+g \rightarrow q+g$ scattering. 
The solid line identifies the quasi-singlet eigenvalue, $\hat{\lambda}_1$,
the dashed and short dashed lines the two quasi-octets, $\hat{\lambda}_2$ and 
$\hat{\lambda}_3$.\label{eigenplotqgqg}}
\end{center}
\end{figure}

The eigenvectors corresponding to the eigenvalues of Eq.\ (\ref{qgqgeigval}) 
can be written in unnormalized form as                    
\beqa
\hspace{-1cm}e_1&=&\left(\begin{array}{c}
{2i\pi \over 4 {\Delta y}-2i\pi+
e^{\frac{2i\pi}{3}}\left( {s_1}+ {s_2}
\right)} \vspace{0.2cm}\\
\frac{3i\pi}{-2 {\Delta y}+i\pi+
e^{\frac{2i\pi}{3}}\left( {s_1}+ {s_2}
\right)} \vspace{0.2cm}\\
1 \end{array} \right) \, \; , \quad
e_2=\left(\begin{array}{c}
\frac{2i\pi}{4 {\Delta y}-2i\pi+\left(e^{\frac{4i\pi}{3}} {s_1}+ {s_2}\right)} \vspace{0.2cm}\\
\frac{3i\pi}{-2 {\Delta y}+i\pi+
\left(e^{\frac{4i\pi}{3}} {s_1}+ {s_2}
\right)} \vspace{0.2cm}\\
1 \end{array}\right)  \nonumber \\ 
&&\nonumber\\
e_3&=&\left(\begin{array}{c}
\frac{2i\pi}{4 {\Delta y}-2i\pi+\left( {s_1}+
e^{\frac{4i\pi}{3}} {s_2}\right)} \vspace{0.2cm}\\
\frac{3i\pi}{-2 {\Delta y}+i\pi+\left( {s_1}+
e^{\frac{4i\pi}{3}} {s_2}\right)
} \vspace{0.2cm}\\
1 \end{array}\right).
\eeqa
They depend only on the geometrical parameter $\Delta y$. It can be easily checked that 
in the limit $\Delta y \rightarrow \infty$ they become pure color states, $e_1$ a singlet,
$e_2$ and $e_3$ antisymmetric and symmetric combinations of the two color octets in 
the basis, Eq.\ (\ref{qgqgbasis}).
However, for the values of $\Delta y$ we are interested in, they are color-mixed states.
Following Ref.\ \cite{StOd1}, we will refer to $e_1$ as the quasi-singlet, and to $e_2$
and $e_3$ as the two quasi-octets.  
Numerical values of the real and imaginary parts of the eigenvalues are  represented in Fig.\ \ref{eigenplotqgqg}.
According to the discussion at the end of Sec.\ \ref{evosec}, the quasi-octet components
of the hard scattering, for large values of $\Delta y$,
will have strong suppression factors, compared to the 
quasi-singlet, because their real parts
grow much faster. In fact, we
will see in Sec.\ \ref{results} that, in this situation, only the quasi-singlet component
of the scattering survives.

\subsection{Soft anomalous dimension for  $gg \rightarrow gg$} 
We work in the basis of Eq.\ (\ref{ggggbasis}).
The structure of the soft anomalous dimension matrix is block-diagonal:
\beqa
 {\Gamma_S}\left(\Delta \eta,
\Delta y \right)=\left(\begin{array}{cc}
	     {\Gamma_{3 \times 3}} & 0_{3 \times 5} \\
	      0_{5 \times 3}      &  {\Gamma_{5 \times 5}}
\end{array} \right)\, ,
\eeqa
with the matrix $\Gamma_{3 \times 3}$ given by:
\beqa
 {\Gamma_{3 \times 3}}=\frac{\alpha_s}{2\pi}
\left(\begin{array}{ccc}
3\rho_2-3i\pi & 0 & 0\\
0 & -3(\xi_2-\rho_2) &0\\
0 & 0 & 3\rho_2+3i\pi 
\end{array}
\right)\, ,
\eeqa 
and the matrix $\Gamma_{5 \times 5}$ given by:
\beqa
{\Gamma_{5 \times 5}}=\frac{\alpha_s}{2\pi}
\left(\begin{array}{ccccc}
3\xi_2+\rho_2+2i\pi & 0 & -12i \pi 
& 0 & 0\\
0 & 3\rho_2 & -3i\pi & -6 i \pi &0 \\
-\frac{3}{2}i\pi& -3i\pi & 3\rho_2 &0 &-\frac{9}{2}
i\pi \vspace{0.3cm} \\ 
0 & -\frac{12}{5}i \pi & 0 & 
-3(\xi_2-\rho_2)& -\frac{18}{5}i\pi \\
0 & 0 & -\frac{4}{3} i \pi & -\frac{8}{3} i\pi &3\rho_2-
5\xi_2
\end{array}
\right).
\eeqa  
The functions $\xi_2$ and $\rho_2$ are very similar to $\xi_1$ and $\rho_1$ 
of Eq.\ (\ref{xi1rho1}),
\beqa 
&&\hspace{-1cm} \g_2(
 {\Delta y})=
i\pi-2 {\Delta y},   \\ 
&&\hspace{-1cm}\w_2(\Delta \eta ,
 {\Delta y})=2\ln \left( \frac{\tanh(\frac{\Delta \eta}{2})+
\tanh\left( \frac{ {\Delta y}}{2}\right)}
{\tanh(\frac{\Delta \eta}{2})-\tanh\left( 
\frac{ {\Delta y}}{2} \right)}\right) 
+3i\pi \, .
\label{xi2rho2}
\eeqa
The eigenvalues of the matrix are
\beqa
\lambda_1&=&\lambda_5=3\frac{\alpha_s}{2\pi} \left( {\rho_2}
-i\pi\right)\nonumber \\
\lambda_2&=&\lambda_4=3\frac{\alpha_s}{2\pi} \left
( {\rho_2}- {\xi_2}\right)\nonumber \\
\lambda_3&=&\lambda_6=3\frac{\alpha_s}{2\pi} \left( {\rho_2}
+i\pi\right)\nonumber \\
\lambda_7&=&\frac{\alpha_s}{2\pi} \left(3
\rho_2- {\xi_2}-4 {\eta_2}\right)\nonumber \\
\lambda_8&=&\frac{\alpha_s}{2\pi} \left(3
{\rho_2}- {\xi_2}+4 {\eta_2}
\right)\, ,
\label{ggggeigval}
\eeqa
where we have introduced the following function of the interjet rapidity: 
\beq
{\eta_2}( {\Delta y})=2\sqrt{ {\Delta y}^2-i\pi
 {\Delta y}-\pi^2}\,.
\label{eta2}
\eeq
Due to the block structure of the soft anomalous dimension matrix, 
the eigenvectors corresponding to the first three eigenvalues
coincide with $c_1$, $c_2$ and $c_3$ of Eq.\ (\ref{ggggbasis}),
while the remaining eigenvectors have zero components along
these color directions. Correspondingly, the structure of
the matrix ${R^{(\a)}}^{-1}$, and consequently $R^{(\a)}$, of Eqs.\ (\ref{eq:newbasS}) and (\ref{newbasH}),
is block-diagonal,
\beq
{R^{(\a)}}^{-1}=\left(\begin{array}{cc}
	      1_{3 \times 3}        & 0_{3 \times 5} \\
	      0_{5 \times 3}      &  {R^{(\a)}}^{-1}_{5 \times 5}
\end{array} \right)\, .
\label{Rblock}
\eeq
Referring to the discussion below  Eq.\ (\ref{ggggsoft}), 
and to Eqs.\ (\ref{eq:newbasS}) and (\ref{newbasH}), we see that  
the dimensionality of the
hard scattering remains $5 \times 5$ after the change from the original basis to 
the basis of the eigenvectors of $\Gamma_S$, because the soft dynamics of 
the physical and unphysical color modes are completely decoupled. 
Here, once again, the eigenvectors depend only on the gap width $\Delta y$.
Explicitly, we find
\beqa
e_i=\left(\begin{array}{c}
      \delta_{i1} \\ 
     \delta_{i2} \\  
	\delta_{i3} \\ 
       0^{(5)} 
\end{array}\right), \; i=1,2,3 \, , \; \; \; 
e_i=\left(\begin{array}{c}
     0^{(3)} \\ 
     e_i^{(5)} 
\end{array}\right), \; i=4 \ldots 8 \, . 
\label{eigstruct}
\eeqa
For $e_4^{(5)}$, $e_5^{(5)}$ and $e_6^{(5)}$
we have the simple results
\beqa
\hspace{-.5cm} e_4^{(5)}=\left(\begin{array}{c}
      0 \vspace{2mm} \\ 
     -\frac{3}{2} \vspace{2mm} \\
      0\vspace{2mm} \\ 
	-\frac{3}{4}-\frac{3i  {\Delta y}}{2\pi} 
\vspace{2mm} \\ 
	1 \end{array} \right), \; \; \;
e_5^{(5)}=\left(\begin{array}{c}
      -15   \vspace{2mm} \\
      \frac{15i}{4\pi}\left(\frac{2i\pi}{5}-2 {\Delta y}
\right) \vspace{2mm} \\
       \frac{15i}{4\pi}\left(2i\pi-2 {\Delta y}
\right)  \vspace{2mm} \\
      3 \vspace{2mm}  \\
1 \end{array} \right), \; \; \; 
e_6^{(5)}=\left(\begin{array}{c}
       -15   \vspace{2mm} \\
      6+\frac{15i}{2\pi} {\Delta y} \vspace{2mm} \\
       -\frac{15i}{2\pi} {\Delta y} \vspace{2mm} \\
      -3 \vspace{2mm}  \\
1 \end{array} \right).
\eeqa
It is easy to see that for large interjet regions ($\Delta y \rightarrow \infty$)
$e_4^{(5)}$ is oriented along the $10 \times 10$ color component, while 
$e_5^{(5)}$ and  $e_6^{(5)}$ are respectively symmetric and antisymmetric combinations
of the $8_S$ and $8_A$ color components. 
\begin{figure}[t]
\begin{center}
\mbox{\epsfysize=6cm \epsfxsize=15cm \epsfbox{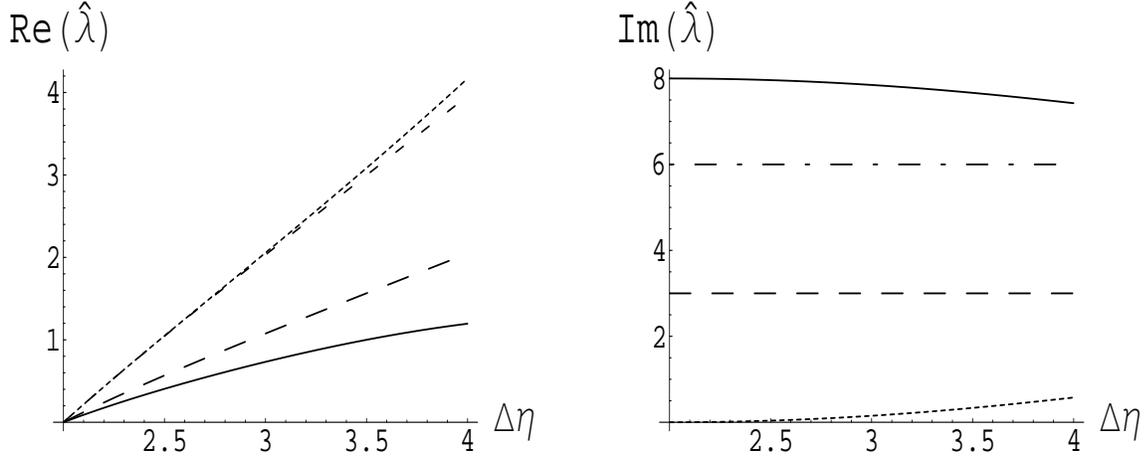}}
\vspace{.3cm}
\caption[dum]{Plot of the real (left) and imaginary (right) part of the eigenvalues ($\hat{\lambda}_i=\lambda_i/\alpha_s$)
of the soft anomalous dimension matrix for $g+g \rightarrow g+g$ scattering. 
In both plots the solid line identifies the quasi-singlet eigenvalue, $\hat{\lambda}_7$,
and the dotted line the quasi-27, $\hat{\lambda}_8$. In the plot of the real parts the dashed
line identifies the two degenerate quasi-octets, $\hat{\lambda}_5$ and $\hat{\lambda}_6$, and the short
dashed line the quasi $10+{\overline {10}}$. In the plot of the imaginary parts the dashed 
line corresponds to both the quasi-octet $\hat{\lambda}_5$ and  the quasi $10+{\overline {10}}$, whereas
the dot dashed line identifies the other quasi-octet, $\hat{\lambda}_6$.\label{eigenplotgggg}}
\end{center}
\end{figure}
The last two eigenvectors have a more complicated analytic form:
\beqa
e_7^{(5)}&=&e_+^{(5)} \nonumber\\
e_8^{(5)}&=&e_-^{(5)} \, ,
\eeqa
with:
\beqa
  \left(e_{\pm}^{(5)}\right)_1&=&\frac{27}{{\cal{N}}_{\pm} \pi^2}\left[32  {\Delta y}^4
-16i {\Delta y}^3\left(4\pi \pm 
i {\eta_2}\right)-4\pi {\Delta y}^2\left(19\pi \pm 6i  {\eta_2}\right)
 \right. \nonumber \\
&&\left.  
+4\pi^2 i {\Delta y}
\left(11\pi \pm 5i {\eta_2}\right) +\pi^3\left(13\pi \pm 6i
 {\eta_2}\right) \right]
\nonumber \\
\left(e_{\pm}^{(5)}\right)_2&=&\frac{27}{{\cal{N}}_{\pm}}
\left[-4 {\Delta y}^2-2 {\Delta y}\left(
-2i\pi \pm  {\eta_2}\right) + \pi\left(3\pi \pm 
 i{\eta_2}\right) \right]
\nonumber \\ 
\left(e_{\pm}^{(5)}\right)_3&=&\frac{9}{{\cal{N}}_{\pm}\pi} \left[24i 
{\Delta y}^3+12
 {\Delta y}^2\left(3\pi \pm i {\eta_2}
\right) +2\pi {\Delta y}\left(\pm 6 {\eta_2}
-17i\pi\right)\right. \nonumber\\
&& \left. -\pi^2 \left(11\pi \pm 7i {\eta_2}\right) \right] \nonumber \\
\left(e_{\pm}^{(5)}\right)_4&=&
\frac{-9\pi}{{\cal{N}}_{\pm}} \left(2i {\Delta y}+\pi\pm2i {\eta_2}\right)
\nonumber \\ 
\hspace{-2cm}\left(e_{\pm}^{(5)}\right)_5&=&1 \, .
\eeqa   
In the denominators, the normalization ${\cal{N}}_{\pm}$ is given by
\beq
{\cal{N}}_{\pm}=
{-28 {\Delta y}^2
+4 i \Delta y \left(7 \pi \pm i \eta_2 \right)
+\pi\left(31\pi\pm 2i {\eta_2}\right)} \, .
\eeq
It is easily  checked that $e_7^{(5)}$, in the limit of a very wide interjet region,
reduces to a pure color singlet state, while $e_8^{(5)}$ points
to the $27$ color direction. Again, for typical values of $\Delta y$,
these eigenvectors are not pure color states, but rather reflect
mixing between the different color components of the hard scattering.
From the eigenvalues of Eq.\ (\ref{ggggeigval}), plotted in Fig.\ \ref{eigenplotgggg}, 
we can see, as already for $q+g \rightarrow q+g$, that
the quasi-singlet component of the scattering, $e_7^{(5)}$, 
dominates the others in the limit of large gaps.

\subsection{Soft anomalous dimension for $q\bar{q} \rightarrow q\bar{q}$}
In the basis of Eq.\ (\ref{qqbbasis}), the soft anomalous dimension
matrix is
\beqa
 {\Gamma_S\left(\Delta \eta,
\Delta y \right)=\frac{\alpha_s}{4\pi} \left(
\begin{array}{cc}
\w_0+ \g_0 & -4\frac{C_F}{N_c}i\pi \\
-8i\pi & \w_0-\g_0
\end{array} \right)},
\label{rapanodim}
\eeqa
where the functions $\g_0$ and $\w_0$ are the  analogs of those
in Eqs. (\ref{xi1rho1}) and 
(\ref{xi2rho2}), 
\beqa 
&&\hspace{-1cm} \g_0( {\Delta y})=
-2N_c  {\Delta y} +2i\pi\frac{N_c^2-2}{N_c}  \, , \nonumber \\
&&\hspace{-1cm} \w_0\left(\Delta \eta,
\Delta y \right)=2\frac{N_c^2-1}{N_c} \ln \left( \frac{\tanh\left(\frac{\Delta \eta}{2}\right)+
\tanh\left( \frac{ {\Delta y}}{2}\right)}
{\tanh\left(\frac{\Delta \eta}{2}\right)-\tanh\left( \frac{ {\Delta y}}{2}\right)}\right) 
\nonumber \\ 
&&\hspace{1.5cm}
+\frac{2}{N_c} {\Delta y}-2i\pi\frac{N_c^2-2}{N_c}\, .
\label{defntw}
\eeqa
In particular, $\rho_0$, which depends on the scattering angle,
differs slightly from the corresponding function computed in Ref.\ \cite{StOd1} for the Tevatron
gaps, because of
the different underlying kinematics. On the other hand $\xi_0$, depending only
on the geometry, through $\Delta y$, is identical. 
\begin{figure}[t]
\begin{center}
\mbox{\epsfysize=6cm \epsfxsize=15cm \epsfbox{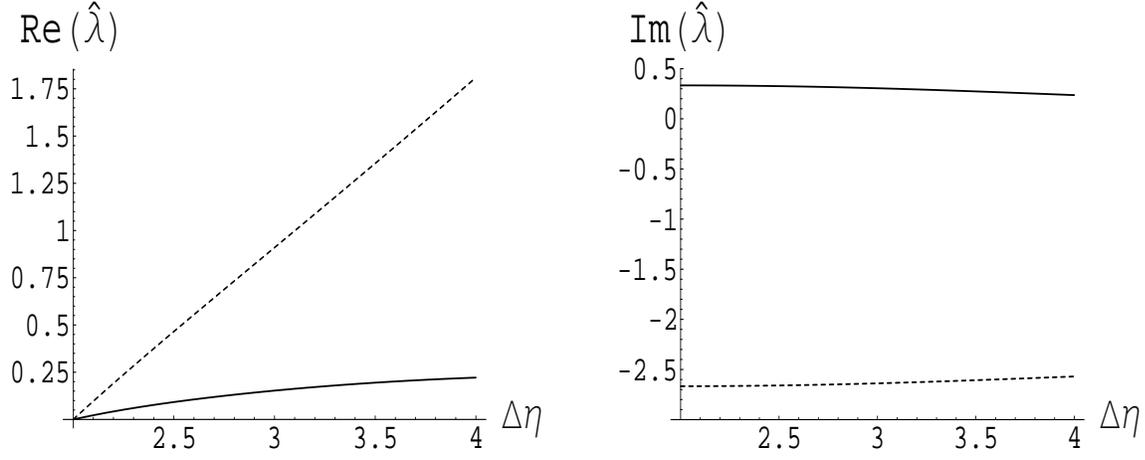}}
\vspace{.3cm}
\caption[dum]{Plot of the real (left) and imaginary (right) part of the eigenvalues ($\hat{\lambda}_i=\lambda_i/\alpha_s$)
of the soft anomalous dimension matrix for $q+\bar{q} \rightarrow q+\bar{q}$ scattering. 
The solid line identifies the quasi-singlet eigenvalue, $\hat{\lambda}_1$,
the short dashed line the quasi-octet, $\hat{\lambda}_2$.\label{eigenplotqqbqqb}}
\end{center}
\end{figure}
The two eigenvalues are
\beqa
 {\lambda_1}\left(\Delta \eta,
\Delta y \right)&=&\frac{\alpha_s}{2\pi}\left[\frac{1}{2}\, \w_0-
\frac{1}{2\sqrt{N_c}} \sqrt{N_c 
\g_0^2 -32C_F\pi^2} \right] 
\nonumber \\
 {\lambda_2}\left(\Delta \eta,
\Delta y \right)&=&\frac{\alpha_s}{2\pi}\left[\frac{1}{2}\w_0+
\frac{1}{2\sqrt{N_c}} \sqrt{N_c 
\g_0^2 -32C_F\pi^2} \right] ,
\label{eigenvalues}
\eeqa
corresponding to the eigenvectors
\beqa
e_1&=&\left(\begin{array}{c}
 {1}\\
 {\frac{8\pi}{i}\left(\g_0-\frac{1}{\sqrt{N_c}}\sqrt{N_c \left(
\g_0(\Delta y)\right)^2 -32C_F\pi^2} \right)^{-1} }\\
\end{array} \right) \nonumber \\ && \nonumber \\
e_2&=&\left(\begin{array}{c}
 {\frac{i}{8\pi}\left(\g_0+\frac{1}{\sqrt{N_c}}\sqrt{N_c \left(
\g_0(\Delta y)\right)^2 -32C_F\pi^2} \right)} \\
 {1}\\
\end{array} \right) \, ,
\label{eigenvectors}
\eeqa
which coincide precisely with the ones given in Ref.\ \cite{StOd1},
since they depend only on the geometric parameter $\Delta y$. 
In the limit of an asymptotically large gap, $e_1$ becomes a pure color singlet
vector, while $e_2$ orients itself along the octet direction.
Referring to $e_1$ and $e_2$ as quasi-singlet and quasi-octet, and looking
at the behavior of the real parts of the eigenvalues, Fig.\ \ref{eigenplotqqbqqb}, we can predict
that, also for this partonic process, the quasi-singlet 
component of the scattering in Eq.\ (\ref{factor2}) will dominate
the quasi-octet in the large gap limit.

\subsection{Soft anomalous dimension for $qq \rightarrow qq$}
In the basis of
Eq.\ (\ref{qqbbasis}), the soft anomalous dimension matrix is
\beqa
 {\Gamma_S\left(\Delta \eta,
\Delta y \right)=\frac{\alpha_s}{4\pi} \left(
\begin{array}{cc}
\w'_0+\g'_0-\frac{4 i\pi}{N_c}  & 4\frac{C_F}{N_c}i\pi \\
8i\pi & \w'_0- \g'_0-\frac{4 i\pi}{N_c}
\end{array} \right)},
\label{qqrapanodim}
\eeqa
where the functions $\w'_0$ and $\g'_0$ are:
\beqa 
&&\hspace{-1cm} \g'_0(\Delta y)=-2N_c  \Delta y +
\frac{4 i\pi}{N_c} \, , \nonumber \\
&&\hspace{-1cm} \w'_0\left(\Delta \eta,
\Delta y \right)=2 \frac{N_c^2-1}{N_c} \left[ \ln \left( \frac{\tanh\left(\frac{\Delta \eta}{2}\right)+
\tanh\left( \frac{\Delta y}{2}\right)}
{\tanh\left(\frac{\Delta \eta}{2}\right)-\tanh\left( \frac{\Delta y}{2}\right)} \right) +2i\pi \right]
+\frac{2}{N_c} \Delta y \, . 
\nonumber \\ 
\label{qqqqdefntw}
\eeqa
\begin{figure}[t]
\begin{center}
\mbox{\epsfysize=6cm \epsfxsize=15cm \epsfbox{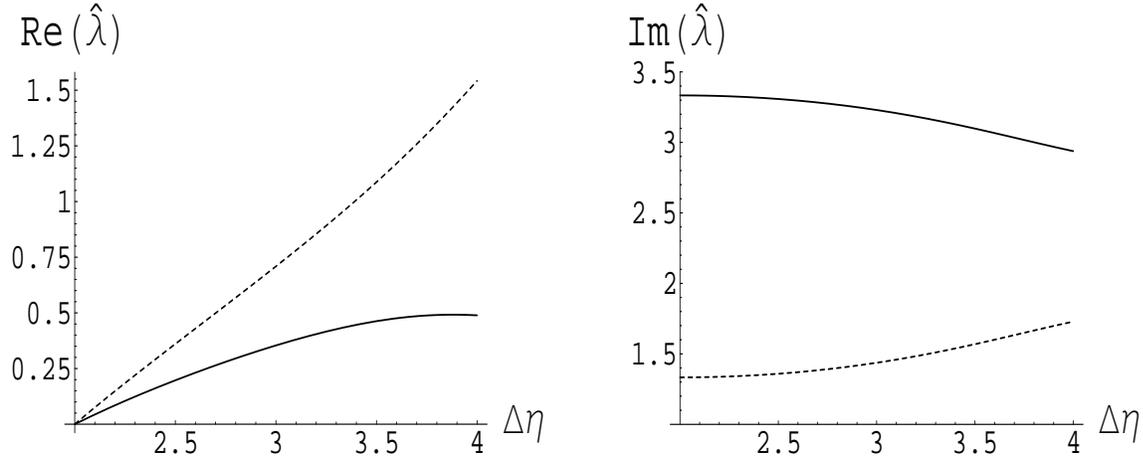}}
\vspace{.3cm}
\caption[dum]{Plot of the real (left) and imaginary (right) part of the eigenvalues ($\hat{\lambda}_i=\lambda_i/\alpha_s$)
of the soft anomalous dimension matrix for $q+q \rightarrow q+q$ scattering. 
The solid line identifies the quasi-singlet eigenvalue, $\hat{\lambda}_1$,
the short dashed line the  quasi-octet, $\hat{\lambda}_2$.\label{eigenplotqqqq}}
\end{center}
\end{figure}
The two eigenvalues are:
\beqa
 {\lambda_1}\left(\Delta \eta,
\Delta y \right)&=&\frac{\alpha_s}{2\pi}\left[\frac{1}{2} \w'_0-
\frac{1}{2\sqrt{N_c}}\sqrt{N_c{\g'_0}^2 -32\pi^2 C_F} -\frac{2 i\pi}{N_c} \right]
\nonumber \\
 {\lambda_2}\left(\Delta \eta,
\Delta y \right)&=&\frac{\alpha_s}{2\pi}\left[\frac{1}{2}\w'_0+
\frac{1}{2\sqrt{N_c}} \sqrt{N_c{\g'_0}^2 -32\pi^2 C_F } -\frac{2 i\pi}{N_c} \right] \, ,
\label{eigenvaluesqq}
\eeqa
which correspond to the eigenvectors:
\beqa
e_1&=&\left(\begin{array}{c}
 \frac{4 \pi C_F}{i N_c} \left(\g'_0+\frac{1}{\sqrt{N_c}}\sqrt{{\g'_0}^2 -32 \pi^2 C_F}\right)^{-1}\\
 1\\
\end{array} \right) \nonumber \\ && \nonumber \\
e_2&=&\left(\begin{array}{c}
 {1}\\
 \frac{i N_c}{4 \pi C_F} \left(\g'_0-\frac{1}{\sqrt{N_c}}\sqrt{{\g'_0}^2 -32\pi^2 C_F}\right)\\
\end{array} \right)  \, .
\label{eigenvectorsqq}
\eeqa
Notice the similarity with the process $q \bar{q} \rightarrow q \bar{q}$.
We can again identify in $e_1$ the quasi-singlet, and in $e_2$ the quasi-octet.
For large gaps, the quasi-octet component of the scattering  will be strongly suppressed with respect to the 
quasi-singlet,
because the real part of the corresponding eigenvalue grows much faster with $\Delta \eta$, as
graphically shown in Fig.\ \ref{eigenplotqqqq}. 

\subsection{Soft anomalous dimension for $\gamma g \rightarrow q\bar{q}$}
\begin{figure}
\begin{center}
\mbox{\epsfysize=6cm \epsfxsize=15cm \epsfbox{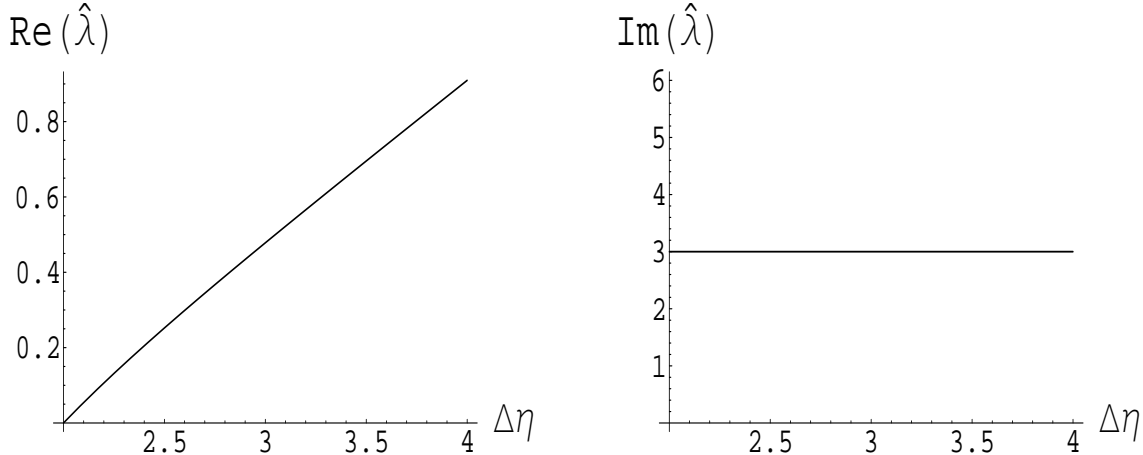}}
\vspace{.3cm}
\caption[dum]{Plot of the real (left) and imaginary (right) part of the 
soft anomalous dimension  for $\gamma+g \rightarrow q+\bar{q}$ scattering.\label{eigenplotdir}}
\end{center}
\end{figure}

For the direct process $\gamma g \rightarrow q\bar{q}$, which, from Fig.\ \ref{figLOpart}, gives the second largest
contribution to the denominator of the fraction,
after the partonic reaction $qg \rightarrow qg$, there is only one possible
color flow at the hard scattering. Thus, the soft anomalous dimension
reduces to a function, given by
\beq
\Gamma_S\left(\Delta \eta,
\Delta y \right)=\frac{\alpha_s}{2\pi}\left[
C_F\ln \left( \frac{\tanh\left(\frac{\Delta \eta}{2}\right)+
\tanh\left( \frac{\Delta y}{2}\right)}
{\tanh\left(\frac{\Delta \eta}{2}\right)-\tanh\left( \frac{\Delta y}{2}\right)} \right) +2i\pi C_A
\right] \, ,
\label{gammadir}
\eeq
whose real and imaginary \footnote{The imaginary part of Eq.\ (\ref{gammadir}) can be extracted
directly from Ref.\ \cite{LOS}.} parts are plotted in Fig.\ \ref{eigenplotdir}. 
We see that the real part of $\Gamma_S$
grows faster with $\Delta \eta$ than the real parts of the quasi-singlet 
eigenvalues for
all the resolved processes, apart from  $gg \rightarrow gg$, Figs.\ 
\ref{eigenplotqgqg}~-~\ref{eigenplotqqqq}. 
In fact, we will see in the next section that the corresponding contribution to the gap cross 
section will have a fast decay rate.

\section{Numerical Results and Discussion}
\label{results}
From the previous two sections, we have all the tools necessary for the evaluation of 
the partonic gap cross section, Eq.\ (\ref{nume}), which we convolute
with the parton distributions and with 
the photon distribution in the electron, according to Eq.\ (\ref{factcs1}), to
get the full gap cross section.
The necessary numerical integrals have been performed with the routine VEGAS.
We have used the CTEQ4L parton distribution set \cite{cteq4} for the proton,
and the GRV-G LO set \cite{GRV} for the photon.
For the interjet energy threshold $Q_0$, which in our
approach defines
the gap event (see Sec.\ \ref{fraction}), we have considered several different values,
from $Q_0=350 {\rm{MeV}}$ up to $Q_0=1050 {\rm{MeV}}$, and we have evaluated the related gap cross section
as a function of the pseudorapidity difference between the jets, $\Delta \eta$.
\begin{figure}[ht]
\begin{center}
\mbox{\epsfysize=12cm \epsfxsize=15cm \epsfbox{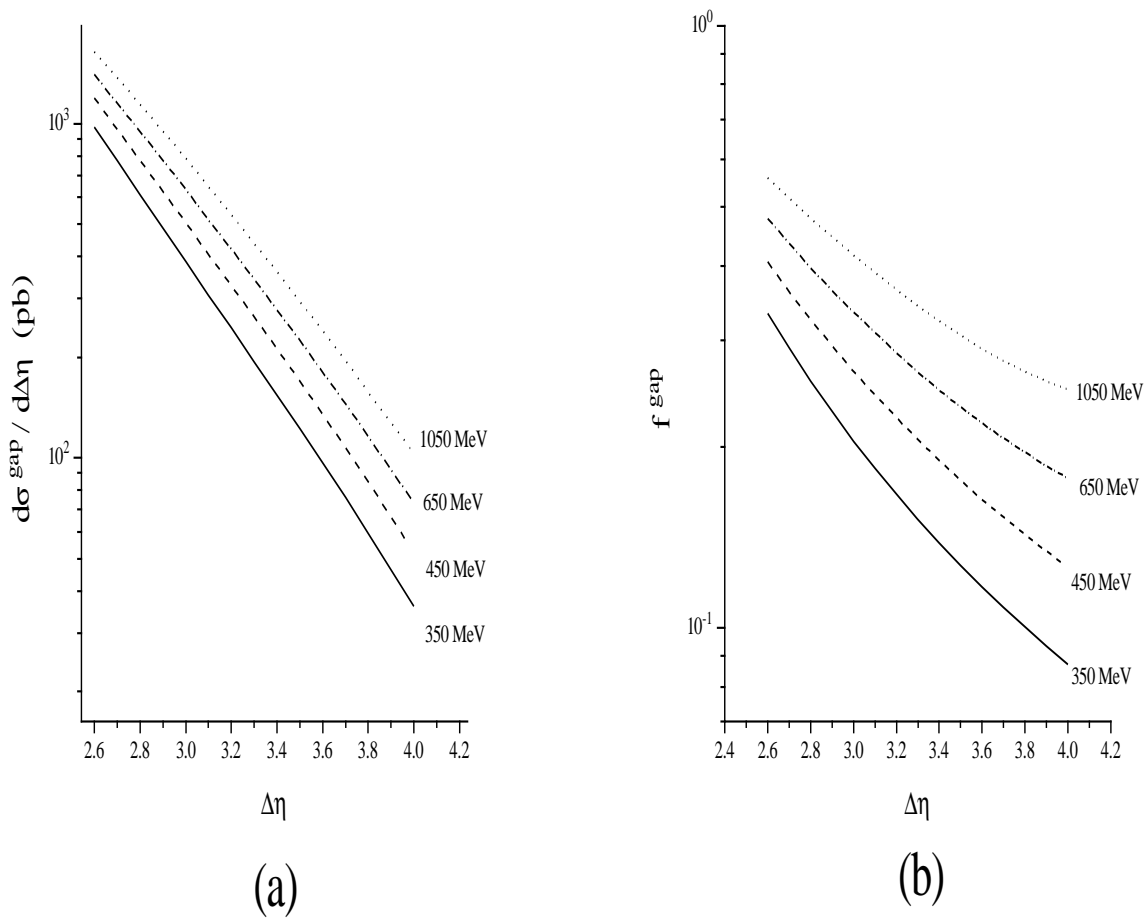}}
\caption[dum]{The gap cross section (a) and fraction (b) at different
values of the energy threshold $Q_0$ identifying the gap 
(see Eq.\ (\ref{nume})).\label{alnum3}}
\end{center}
\end{figure}
Figs.\ \ref{alnum3}a,b show our results for the gap cross section and
the corresponding gap fraction. Both quantities increase with increasing $Q_0$, simply because
a stronger limitation on  interjet radiation leads to a stronger suppression of the result.
For the most forward configuration of the dijet analyzed in Ref.\ \cite{ZEUS}, $\Delta \eta=3.75$,
the gap fraction, Fig.\ \ref{alnum3}b, varies from about $10 \%$ at  $Q_0=350 {\rm{MeV}}$ to about $25 \%$ at 
$Q_0=1050 {\rm{MeV}}$.
One may ask, for both the gap cross section and the gap fraction, whether any of the
plots in Figs.\ \ref{alnum3}a,b fits the experimental data of Ref.\ \cite{ZEUS}, 
for which, as discussed in Sec.\ \ref{fraction}, the gap event is identified in a related, but different way.
\begin{figure}[ht]
\begin{center}
\mbox{\epsfysize=12cm \epsfxsize=15cm \epsfbox{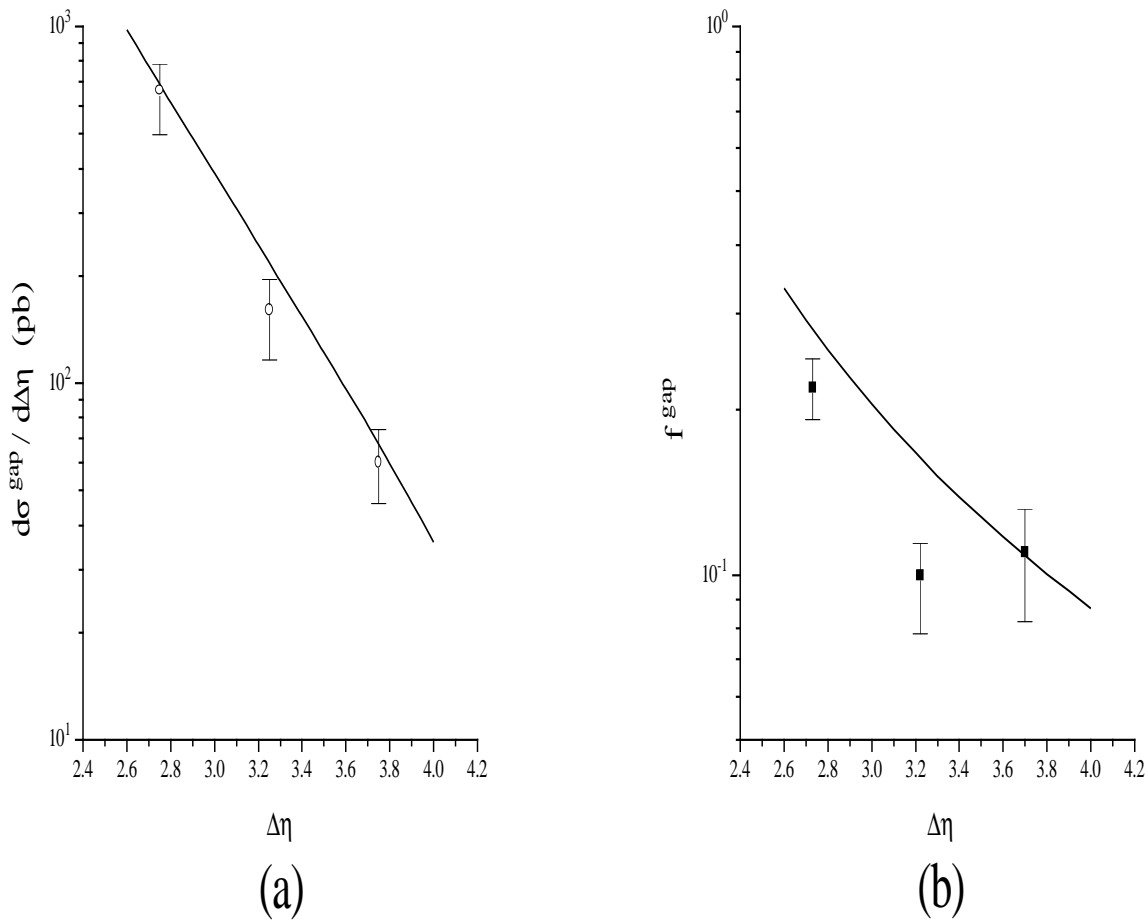}}
\caption[dum]{The gap cross section (a) and fraction (b),  for
the gap identified by the  energy threshold $Q_0=350 {\rm{MeV}}$, compared
with the data of Ref.\ \cite{ZEUS}.\label{alnum2}}
\end{center}
\end{figure}
With the choice $Q_0=350 {\rm{MeV}}$ for the interjet energy, we obtain
a good match, as shown in Fig.\ \ref{alnum2}a,b.
Observe that, in this situation, $Q_0$ is not much larger than $\Lambda$ in Eq.\ (\ref{nume}).
Results in this region, at the boundary with non-perturbative QCD, can be considered
as smooth extrapolations of perturbative QCD resummation.
\begin{figure}[ht]
\begin{center}
\mbox{\epsfysize=8.6cm \epsfxsize=8.6cm \epsfbox{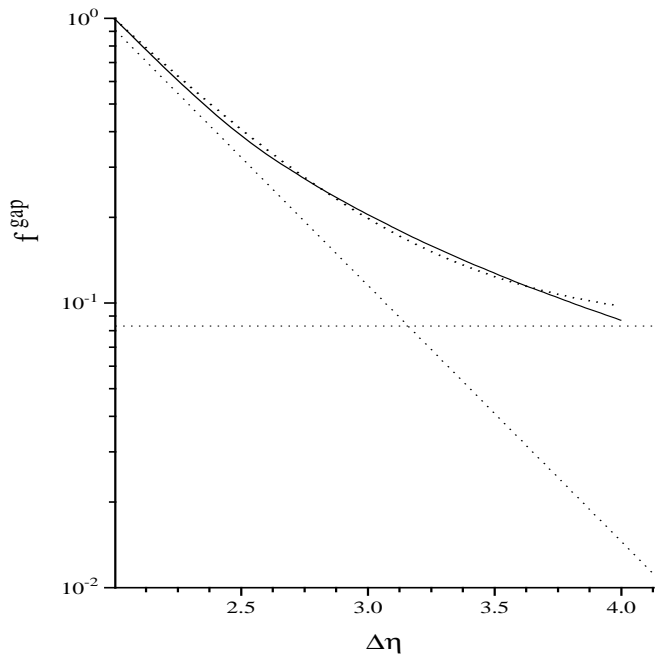}}
\vspace{.3cm}
\caption[dum]{Analog of Fig.\ 3d of Ref.\  \cite{ZEUS}. The gap fraction of Fig.\ \ref{alnum2}
(solid line) is redisplayed and compared with the result of a fit to an exponential plus a constant.\label{alnum1} }
\end{center}
\end{figure}

In Fig.\ \ref{alnum1}, the analog of Fig.\ 3d of Ref.\ \cite{ZEUS},
we show again our gap fraction at $Q_0=350 {\rm{MeV}}$, from Eqs. (\ref{nume}) and (\ref{deno}), 
and compare it with the result of a two parameter fit 
to the expression
\beq
f^{\rm{fit}}(\alpha,\beta,\Delta \eta)=C(\alpha, \beta)\, e^{\alpha \Delta \eta}+\beta .
\label{ffit}
\eeq
as done in Ref.\ \cite{ZEUS}. 
We find $\alpha=-2.1$ and $\beta=8.3 \%$, in good 
agreement with the values of Ref.\ \cite{ZEUS}: $\alpha=-2.7 \pm 0.3 $ and $\beta=(7 \pm 2) \%$.
Here, in analogy with Ref.\ \cite{ZEUS}, we have imposed the constraint 
$f^{\rm{fit}}=1$ at $\Delta \eta=2$, meaning that the gap cross section has
to reduce to the full dijet cross section in the absence of a gap.
Notice that this information is not encoded in Eq.\ (\ref{nume}), because the last two factors on
the right hand side do not reduce exactly to unity for $\Delta \eta =2$.
This can be seen from Eq.\ (\ref{expon}) and from the 
eigenvalues of the soft anomalous dimension matrices, 
Figs.\ \ref{eigenplotqgqg}-\ref{eigenplotdir}. 
In the  limit $\Delta \eta \rightarrow 2$,  
the real parts of the eigenvalues
vanish, but the imaginary parts do not, 
causing a slight suppression  
in the mixed terms, ($\beta \neq \gamma$), of the hard scattering matrix in Eq.\ (\ref{nume}).
The validity of our formula for the gap cross section, Eq.\ (\ref{nume}), is restricted to
interjet rapidity widths $\Delta y$ neither too small, nor too large. 
We will limit ourselves to the $\Delta \eta$ range $2.6 < \Delta \eta < 4.0$.

The reasons for these restrictions are as follows. Our 
expression
in Eq.\ (\ref{nume})
is the result of the resummation of logarithms of the soft interjet energy, 
$\alpha_s^n(-\hat{t}) \ln^n(Q_0/ \sqrt{-\hat{t}})$, which are assumed
to be the only large, perturbatively dangerous quantities in the problem.
When $\Delta y \rightarrow 0$, the gap closes and the cross section becomes inclusive.
More specifically, logarithms of $\sqrt{-\hat{t}}$ would be suppressed by powers of
$\Delta y$. Analogously, when $\Delta y \rightarrow \infty$ (i.\ e.\ ,
 $\Delta \eta \rightarrow \infty$),
terms of the form $\alpha_s^n(-\hat{t})(\Delta y)^n$, 
coming from the eigenvalues 
of the $\Gamma^{(\a)}_S$'s, are large at each order
in perturbation theory, and we need further resummations.
We should also emphasize that  the  limit $\Delta \eta \rightarrow \infty$ 
corresponds to the Regge region of the scattering,
when $\hat{s}$ becomes large at fixed momentum transfer, $\hat{t}$.
Resummation in the Regge limit organizes logarithms of
the form $\alpha_s^n \ln^n\left(\hat{s}/(-\hat{t})\right)$, coming
from the BFKL ladders of gluons \cite{DDucaT}. For double
logarithmic terms like $\alpha_s^n \ln^n\left(\hat{s}/(-\hat{t})\right)
\ln^n(Q_0/ \sqrt{-\hat{t}})$, which occur in Reggeized color octet exchange,
both BFKL resummation and our method give the same result.

Note that the suppression we find here is not double logarithmic in $Q_0$ \cite{MRK}. 
Rather, it comes 
from the exponentiation of single ``soft'' logarithms only, 
as shown in Eq.\ (\ref{eq:resoft}). The underlying physical reason is that 
a soft gluon emitted into the interjet region can never become collinear 
to either of the forward partons from which the jets originate.

\begin{figure}[t]
\begin{center}
\mbox{\epsfysize=8.6cm \epsfxsize=8.6cm \epsfbox{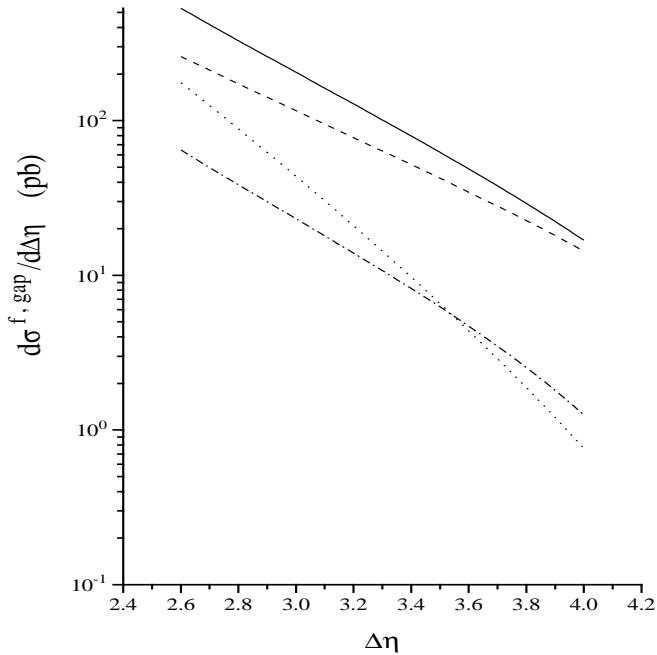}}
\vspace{.3cm}
\caption{Overall contribution to the 
gap cross section from $qg \rightarrow qg$ scattering (solid line). Contribution
from the quasi-singlet component (dashed) and from the two quasi-octet components
(dotted and dot dashed).\label{qg}}
\end{center}
\end{figure}

\begin{figure}[ht]
\begin{center}
\mbox{\epsfysize=8.6cm \epsfxsize=8.6cm \epsfbox{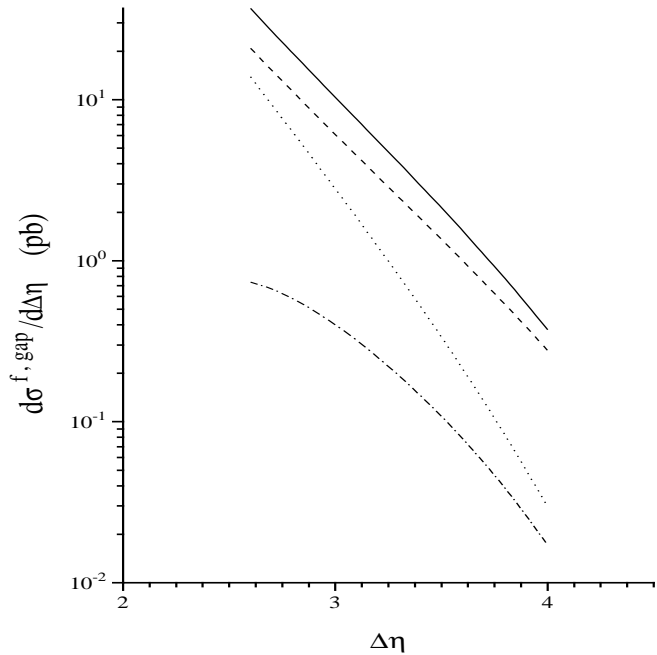}}
\vspace{.3cm}
\caption{Overall contribution to the 
gap cross section from $gg \rightarrow gg$ scattering (solid line). Contribution
from the quasi-singlet component (dashed) and from the two quasi-octet components
(dotted and dot dashed). The contributions of the quasi-$10 \times \overline{10}$
and of the quasi-$27$ component are not exhibited, because they are too small.\label{gg}}
\end{center}
\end{figure}

\begin{figure}[ht]
\begin{center}
\mbox{\epsfysize=8.6cm \epsfxsize=8.6cm \epsfbox{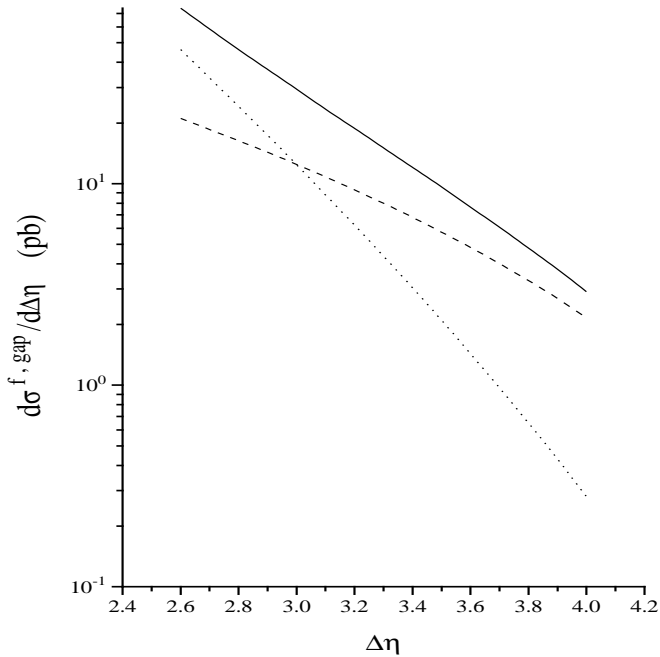}}
\vspace{.3cm}
\caption{Overall contribution to the 
gap cross section from $q\bar{q} \rightarrow q\bar{q}$ scattering (solid line). Contribution
from the quasi-singlet component (dashed) and from the quasi-octet component
(dotted).\label{qqb}}
\end{center}
\end{figure}

\begin{figure}[ht]
\begin{center}
\mbox{\epsfysize=8.6cm \epsfxsize=8.6cm \epsfbox{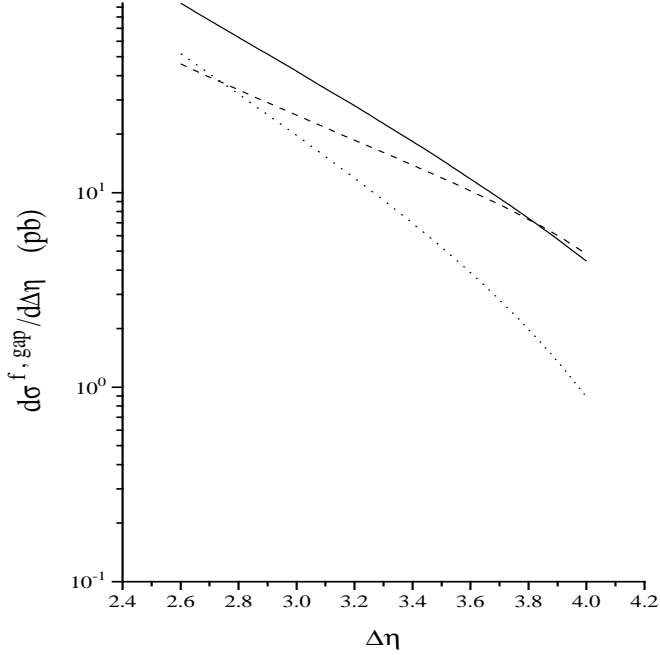}}
\vspace{.3cm}
\caption[dum]{Overall contribution to the 
gap cross section from $q q \rightarrow q q$ scattering (solid line). Contribution
from the quasi-singlet component (dashed) and from the quasi-octet component
(dotted). Negative interference terms \cite{StOd1}, corresponding to
$\beta \neq \gamma$ in Eq.\ (\ref{nume}), are not exhibited separately.
\label{qq}}
\end{center}
\end{figure}

Finally, in Figs.\ \ref{qg}-\ref{qq} we show the 
contributions to the gap cross section, Eq.\ (\ref{nume}) at $Q_0=350 {\rm{MeV}}$, from the 
different partonic processes studied in Sec.\ \ref{anodim}. In each case we give the full result and 
its decomposition into quasi-color components. We see that, as anticipated
in Sec.\ \ref{anodim}, the quasi-singlet components,
in all cases,
give almost the whole gap cross section at large values of $\Delta \eta$.
It then becomes 
clear that the parameter $\beta=8.3 \%$, extracted from the fit, 
Eq.\ (\ref{ffit}),
can be associated with the asymptotic quasi-singlet fraction.
From Figs.\ \ref{eigenplotqgqg}-\ref{eigenplotqqqq} we see that at high $\Delta \eta$ 
the real parts of
the quasi-singlet eigenvalues are ordered,
from the lowest value 
of $q \, \bar{q} \rightarrow q \, \bar{q}$ scattering, 
through the intermediate values
of processes $q \, g \rightarrow q \,  g$
and $q \, q \rightarrow q \,  q$,
up to the largest value 
of $g \, g \rightarrow g \,  g$ scattering.
In Figs.\ \ref{qg}-\ref{qq}, the decay slopes of the
quasi-singlet components of the corresponding contributions
to the gap cross section
reflect the same ordering property,
as a consequence of Eq.\ (\ref{factor2}).  
This observation indicates
that gaps tend to be more easily formed with scattered quarks 
and antiquarks than with gluons \cite{eboli}.

\section{Conclusions}
In this paper, we have shown that it is possible to analyze the dijet rapidity gap events from
photoproduction, observed at HERA, by introducing an energy-dependent  definition of the gap, as already
done in Ref.\ \cite{StOd1} for the Tevatron events.
The result we have found is perturbative, as the ordering of the different scales in Eq.\ (\ref{factor2}) shows, with 
$\Lambda < Q_c < \sqrt{-\hat{t}}$.
The experimental behavior of the gap fraction is approximately reproduced by fixing the threshold 
of interjet energy, which defines the gap, at $Q_0=350 {\rm{MeV}}$.
Our result also predicts how gap fractions increase with $Q_0$. 
We conjecture that in photoproduction non-perturbative ``survival'' effects from the interactions of spectator partons,
which can give rise to interjet multiplicity \cite{GLM}, should be
reduced with respect to the case of $p \bar{p}$ scattering, because 
there is only one incoming hadron.
The experimental determination of 
the double differential cross section $d^2\sigma_{ep}/d\Delta \eta dQ_c$, 
(and related fraction), Eq.\ (\ref{nume}), for different values of the interjet energy flow
identifying the gap,
as in Fig.\ \ref{alnum3}, would offer
a significant test of the perturbative dynamics of QCD radiation.

\subsection*{Acknowledgments}
I would like to express my gratitude to Prof.\ George Sterman,
for many conversations and insights, and for all his help.
I am indebted to Brian Harris and 
Jack Smith for their valuable advice 
in the implementation of the numerical integrations.  
Finally I would like to thank Alan White, Jim Whitmore, and the other participants
of the Theory Institute on Deep-Inelastic Diffraction
(Argonne National Lab, September 14th - 16th, 1998), where
the early stages of this work were presented, for useful discussions.
This work was
supported in part by the National Science Foundation, grant PHY9722101.
\appendix
\section{The Denominator of the Gap Fraction: Partonic Cross Sections and 
Parton Luminosities}  
\label{AppA}
In this Appendix we will summarize for completeness
the Born cross sections, and corresponding  
parton luminosities, $L^{(\a)}$, for all the partonic processes
contributing to the denominator of the gap fraction, Eq.\ (\ref{deno}).
We will consider for the quarks four active flavors, $n_f=4$,  and assume flavor symmetry.
Correspondingly, the quark luminosities will be expressed as functions of 
the proton valence distributions, $\phi_{u/p}$ and $\phi_{d/p}$, and of the
sea distributions for both the proton and the photon, respectively
$\phi_{{\rm{sea}}/p}$ and $\phi_{{\rm{sea}}/ \gamma}$.
The analytic cross sections will always be of the form
\beqa
\frac{d\hat{\sigma}^{(\a)}}
{d\Delta\eta}&=& \left( 2 \cosh^2 \left( \frac{\Delta \eta}{2} \right)\right)^{-1} 
\,\alpha_s^2(\hat{t}) \, \left( \frac{\pi}{2\hat{s}} \right) \, |{\cal{M}}^{(\a)}(\hat{s},\hat{t},\hat{u})|^2 \nonumber \\
&=&\frac{\pi}{2\hat{s}} \left( 2 \cosh^2 \left( \frac{\Delta \eta}{2} \right)\right)^{-1} 
 H^{(\a)}_{IL}\left(\sqrt{-\hat{t}},
\sqrt{\hat{s}},\mu,\alpha_s(\mu^2) \right) \, 
 S^{(\a)}_{LI} \left(\Delta y,\frac{Q_c}{\mu}\right) \, ,
\label{analcs}
\eeqa
where, in both equalities, 
the first factor is the Jacobian $[d\cos(\hat{\theta})/d\Delta \eta]$, and the rest is the partonic
cross section $d\hat{\sigma}^{(\a)}/d\cos(\hat{\theta})$.
The second equality defines the normalization of the hard scattering matrix in Eq.\ (\ref{factor}).
In the list 
below, we will specify the 
invariant matrix elements squared, $|{\cal{M}}^{(\a)}(\hat{s},\hat{t},\hat{u})|^2$
\cite{KK,ESW}.
\begin{itemize}
\item
Processes $q (\bar{q}) g \rightarrow q (\bar{q}) g$, $g q (\bar{q})  \rightarrow g q (\bar{q})$:
\beqa
L^{(\a)}&=&\phi_{g/ \gamma} \left( \phi_{u/p}+\phi_{d/p}+6 \phi_{{\rm{sea}}/p} \right)+8 
\phi_{{\rm{sea}}/ \gamma} \phi_{g/p}
\nonumber \\
|{\cal{M}}^{(\a)}(\hat{s},\hat{t},\hat{u})|^2&=&\frac{\hat{u}^2+\hat{s}^2}{\hat{t}^2}-\frac{4}{9} \, 
\frac{\hat{s}^2+\hat{u}^2}{\hat{s}\hat{u}}\,.
\label{qgqganalcsl}
\eeqa
\item
Processes $q (\bar{q}) g \rightarrow g q (\bar{q}) $, $g q (\bar{q})  \rightarrow q (\bar{q})g $ (observe that they differ from the processes of Eq.\ (\ref{qgqganalcsl})
by the exchange of particles $1$ and $2$ in Eq.\ (\ref{Mandlst}), and, 
correspondingly, of the Mandelstam invariants $\hat{t}$ and $\hat{u}$):
\beqa
L^{(\a)}&=&\phi_{g/ \gamma} \left( \phi_{u/p}+\phi_{d/p}+6 \phi_{{\rm{sea}}/p} \right)+8 
\phi_{{\rm{sea}}/ \gamma} \phi_{g/p}
\nonumber \\
|{\cal{M}}^{(\a)}(\hat{s},\hat{t},\hat{u})|^2&=&\frac{\hat{t}^2+\hat{s}^2}{\hat{u}^2}-\frac{4}{9} \, 
\frac{\hat{s}^2+\hat{t}^2}{\hat{s}\hat{t}}\,.
\label{qggqanalcsl}
\eeqa
\item
Process $gg \rightarrow gg$:
\beqa
L^{(\a)}&=&\phi_{g/ \gamma} \phi_{g/p}
\nonumber \\
|{\cal{M}}^{(\a)}(\hat{s},\hat{t},\hat{u})|^2&=& \frac{9}{2} \left(3-\frac{\hat{s}\hat{u}}{\hat{t}^2}+
\frac{\hat{t}\hat{u}}{\hat{s}^2}+\frac{\hat{s}\hat{t}}{\hat{u}^2} \right)\,.
\label{gggganalcsl}
\eeqa
\item
Process $qq \rightarrow qq$:
\beqa
L^{(\a)}&=&\phi_{{\rm{sea}}/ \gamma}\left( \phi_{u/p}+\phi_{d/p}+6 \phi_{{\rm{sea}}/p} \right)
\nonumber \\
|{\cal{M}}^{(\a)}(\hat{s},\hat{t},\hat{u})|^2&=&\frac{4}{9} \, 
\frac{\hat{s}^2+\hat{u}^2}{\hat{t}^2} +\frac{4}{9} \, 
\frac{\hat{s}^2+\hat{t}^2}{\hat{u}^2} -\frac{8}{27} \, \frac{\hat{s}^2}{\hat{t}\hat{u}} \,.
\label{qqqqanalcsl}
\eeqa
\item
Process $qq' \rightarrow qq'$:
\beqa
L^{(\a)}&=&3 \phi_{{\rm{sea}}/ \gamma}\left( \phi_{u/p}+\phi_{d/p}+6 \phi_{{\rm{sea}}/p} \right)
\nonumber \\
|{\cal{M}}^{(\a)}(\hat{s},\hat{t},\hat{u})|^2&=&\frac{4}{9} \, 
\frac{\hat{s}^2+\hat{u}^2}{\hat{t}^2} \,.
\label{qqpqqpanalcsl}
\eeqa
\item
Process $qq' \rightarrow q'q$:
\beqa
L^{(\a)}&=&3 \phi_{{\rm{sea}}/ \gamma}\left( \phi_{u/p}+\phi_{d/p}+6 \phi_{{\rm{sea}}/p} \right)
\nonumber \\
|{\cal{M}}^{(\a)}(\hat{s},\hat{t},\hat{u})|^2&=&\frac{4}{9} \, 
\frac{\hat{s}^2+\hat{t}^2}{\hat{u}^2} \,.
\label{qqpqpqanalcsl}
\eeqa
\item
Processes $q\bar{q} \rightarrow q\bar{q}$, $\bar{q}q \rightarrow \bar{q}q$:
\beqa
L^{(\a)}&=&\phi_{{\rm{sea}}/ \gamma}\left( \phi_{u/p}+\phi_{d/p}+6 \phi_{{\rm{sea}}/p} \right)
\nonumber \\
|{\cal{M}}^{(\a)}(\hat{s},\hat{t},\hat{u})|^2&=&\frac{4}{9} \, 
\frac{\hat{s}^2+\hat{u}^2}{\hat{t}^2} +\frac{4}{9} \, 
\frac{\hat{t}^2+\hat{u}^2}{\hat{s}^2} -\frac{8}{27} \, \frac{\hat{u}^2}{\hat{s}\hat{t}} \,.
\label{qqbqqbanalcsl}
\eeqa
\item
Processes $q\bar{q}' \rightarrow q\bar{q}'$, $\bar{q}q' \rightarrow \bar{q}q'$:
\beqa
L^{(\a)}&=&3 \phi_{{\rm{sea}}/ \gamma}\left( \phi_{u/p}+\phi_{d/p}+6 \phi_{{\rm{sea}}/p} \right)
\nonumber \\
|{\cal{M}}^{(\a)}(\hat{s},\hat{t},\hat{u})|^2&=&\frac{4}{9} \, 
\frac{\hat{s}^2+\hat{u}^2}{\hat{t}^2}\,. 
\label{qqbpqqbpanalcsl}
\eeqa
\item
Processes $q\bar{q} \rightarrow q' \bar{q}'$, $\bar{q}q \rightarrow \bar{q}'q'$:
\beqa
L^{(\a)}&=&3 \phi_{{\rm{sea}}/ \gamma}\left( \phi_{u/p}+\phi_{d/p}+6 \phi_{{\rm{sea}}/p} \right)
\nonumber \\
|{\cal{M}}^{(\a)}(\hat{s},\hat{t},\hat{u})|^2&=&\frac{4}{9} \, 
\frac{\hat{t}^2+\hat{u}^2}{\hat{s}^2}\,.
\label{qqbqpqbpanalcsl}
\eeqa
\item
Processes $q\bar{q} \rightarrow \bar{q}q$, $\bar{q}q \rightarrow q\bar{q}$:
\beqa
L^{(\a)}&=&\phi_{{\rm{sea}}/ \gamma}\left( \phi_{u/p}+\phi_{d/p}+6 \phi_{{\rm{sea}}/p} \right)
\nonumber \\
|{\cal{M}}^{(\a)}(\hat{s},\hat{t},\hat{u})|^2&=&\frac{4}{9} \, 
\frac{\hat{s}^2+\hat{t}^2}{\hat{u}^2} +\frac{4}{9} \, 
\frac{\hat{t}^2+\hat{u}^2}{\hat{s}^2} -\frac{8}{27} \, \frac{\hat{t}^2}{\hat{s}\hat{u}} \,.
\label{qqbqbqanalcsl}
\eeqa
\item
Processes $q\bar{q}' \rightarrow \bar{q}'q$, $\bar{q}q' \rightarrow q'\bar{q}$:
\beqa
L^{(\a)}&=&3 \phi_{{\rm{sea}}/ \gamma}\left( \phi_{u/p}+\phi_{d/p}+6 \phi_{{\rm{sea}}/p} \right)
\nonumber \\
|{\cal{M}}^{(\a)}(\hat{s},\hat{t},\hat{u})|^2&=&\frac{4}{9} \, 
\frac{\hat{s}^2+\hat{t}^2}{\hat{u}^2}\,. 
\label{qqbpqbpqanalcsl}
\eeqa
\item
Processes $q\bar{q} \rightarrow \bar{q}'q' $, $\bar{q}q \rightarrow q'\bar{q}'$:
\beqa
L^{(\a)}&=&3 \phi_{{\rm{sea}}/ \gamma}\left( \phi_{u/p}+\phi_{d/p}+6 \phi_{{\rm{sea}}/p} \right)
\nonumber \\
|{\cal{M}}^{(\a)}(\hat{s},\hat{t},\hat{u})|^2&=&\frac{4}{9} \, 
\frac{\hat{t}^2+\hat{u}^2}{\hat{s}^2} \,.
\label{qqbqbpqpanalcsl}
\eeqa
\item
Processes $q\bar{q} \rightarrow gg$, $\bar{q}q \rightarrow gg$
\beqa
L^{(\a)}&=&\phi_{{\rm{sea}}/ \gamma}\left( \phi_{u/p}+\phi_{d/p}+6 \phi_{{\rm{sea}}/p} \right)
\nonumber \\
|{\cal{M}}^{(\a)}(\hat{s},\hat{t},\hat{u})|^2&=&-\frac{8}{3} \, 
\frac{\hat{t}^2+\hat{u}^2}{\hat{s}^2}+\frac{32}{27} \, 
\frac{\hat{t}^2+\hat{u}^2}{\hat{t}\hat{u}}\,.
\label{qqbgganalcsl}
\eeqa
\item
Processes $gg \rightarrow q\bar{q}$, $gg \rightarrow \bar{q}q $
\beqa
L^{(\a)}&=&\phi_{g/ \gamma} \phi_{g/p}
\nonumber \\
|{\cal{M}}^{(\a)}(\hat{s},\hat{t},\hat{u})|^2&=&-\frac{3}{8} \, 
\frac{\hat{t}^2+\hat{u}^2}{\hat{s}^2}+\frac{1}{6} \, 
\frac{\hat{t}^2+\hat{u}^2}{\hat{t}\hat{u}}\,.
\label{ggqqbanalcsl}
\eeqa
\item
Processes $\gamma g \rightarrow q\bar{q}$, $\gamma g \rightarrow \bar{q}q $
\beqa
L^{(\a)}&=&2 \phi_{g/p}
\nonumber \\
|{\cal{M}}^{(\a)}(\hat{s},\hat{t},\hat{u})|^2&=&\frac{10}{9} \, \frac{\alpha_{{\rm{em}}}}{\alpha_s}\, 
\left(\frac{\hat{u}}{\hat{t}}+\frac{\hat{t}}{\hat{u}} \right)\,.
\label{dir1analcsl}
\eeqa
\item
Process $\gamma q(\bar{q}) \rightarrow q(\bar{q}) g$, 
\beqa
L^{(\a)}&=& \frac{4}{9} \phi_{u/p}+\frac{1}{9} \phi_{d/p}+\frac{15}{9} \phi_{{\rm{sea}}/p}
\nonumber \\
|{\cal{M}}^{(\a)}(\hat{s},\hat{t},\hat{u})|^2&=&\frac{\alpha_{{\rm{em}}}}{\alpha_s}\, 
\left(\frac{-\hat{u}}{\hat{s}}+\frac{\hat{s}}{-\hat{u}} \right)\,. 
\label{dir2analcsl}
\eeqa
\end{itemize}

\end{document}